\documentclass[aip,preprint]{revtex4-1}
\usepackage{braket}
\usepackage{graphicx}
\usepackage{color}
\draft

\begin{document}
\makeatletter 
\renewcommand{\figurename}{\textbf{Figure}}
\renewcommand\thefootnote{\arabic{footnote})}
\makeatother

\title{Picosecond coherent electron motion in a silicon single-electron source}

\author{Gento Yamahata\footnote{Electronic mail: gento.yamahata.eu@hco.ntt.co.jp}}
\thanks{These authors contributed equally to this work.}
\affiliation{NTT Basic Research Laboratories, NTT Corporation, 3-1 Morinosato Wakamiya, Atsugi, Kanagawa 243-0198, Japan}

\author{Sungguen Ryu}
\thanks{These authors contributed equally to this work.}
\affiliation{Department of Physics, Korea Advanced Institute of Science and Technology, Daejeon 34141, Korea}

\author{Nathan Johnson}
\affiliation{NTT Basic Research Laboratories, NTT Corporation, 3-1 Morinosato Wakamiya, Atsugi, Kanagawa 243-0198, Japan}

\author{H.-S. Sim} \email{hssim@kaist.ac.kr}
\affiliation{Department of Physics, Korea Advanced Institute of Science and Technology, Daejeon 34141, Korea}

\author{Akira Fujiwara}
\affiliation{NTT Basic Research Laboratories, NTT Corporation, 3-1 Morinosato Wakamiya, Atsugi, Kanagawa 243-0198, Japan}

\author{Masaya Kataoka} \email{masaya.kataoka@npl.co.uk}
\affiliation{National Physical Laboratory, Hampton Road, Teddington, Middlesex TW11 0LW, United Kingdom}

\maketitle
{\bf Understanding ultrafast coherent electron dynamics is necessary for application of a single-electron source to metrological standards\cite{pekola-rev}, quantum information processing\cite{SEsouce1}, including electron quantum optics\cite{Feve_opt1}, and quantum sensing\cite{Nathan_samp, qsen1}. While the dynamics of an electron emitted from the source has been extensively studied\cite{kataokaSAW, Feve_sci_coh, lev-opt,glat_tomo,  Ubbelohde1, Nathan1}, there is as yet no study of the dynamics inside the source. This is because the speed of the internal dynamics is typically higher than 100 GHz, beyond state-of-the-art experimental bandwidth\cite{SEsouce1}. Here, we theoretically and experimentally demonstrate that the internal dynamics in a silicon single-electron source comprising a dynamic quantum dot can be detected, utilising a resonant level with which the dynamics is read out as gate-dependent current oscillations. Our experimental observation and simulation with realistic parameters show that an electron wave packet spatially oscillates quantum-coherently at $\sim$ 200 GHz inside the source. Our results will lead to a protocol for detecting such fast dynamics in a cavity and offer a means of engineering electron wave packets\cite{Sungguen1}. This could allow high-accuracy current sources\cite{gib1, PTB-ulca1, NPL-NTT1, Zhao_pump},
high-resolution and high-speed electromagnetic-field sensing\cite{Nathan_samp}, and high-fidelity initialisation of flying qubits\cite{yamamoto_fly, MeunSAW1}.}

Owing to recent demonstrations of high-accuracy GHz operation\cite{gib1, PTB-ulca1, NPL-NTT1, Zhao_pump}, a single-electron pump with a tunable-barrier quantum dot (QD) becomes promising for application to on-demand single-electron sources\cite{pekola-rev}. Because of the fast dynamic movement of the QD, there can occur nontrivial electron dynamics, such as non-adiabatic excitation\cite{kataoka1} and subsequent coherent time evolution. While the non-adiabatic excitation could degrade the pumping accuracy, a spatial movement of an electron wave packet due to the coherent time evolution can be used for engineering a wave packet emitted from the QD\cite{Sungguen1}, which could make possible electron quantum optics experiments and high-speed quantum sensing with high resolution. In addition, understanding of the fast electron dynamics could offer insight into quantum computing with QDs. However, the existing standard measurement technique\cite{ntte, Petersson1, DKim1} does not have enough bandwidth to detect fast dynamics beyond 100 GHz. In order to overcome the limitation and detect the fast dynamics in the QD, we use a temporal change in a tunnel rate between a resonant level in a tunnel barrier and a QD, which is induced by the dynamic change of the QD potential.

\begin{figure*}
\begin{center}
\includegraphics[pagebox=artbox]{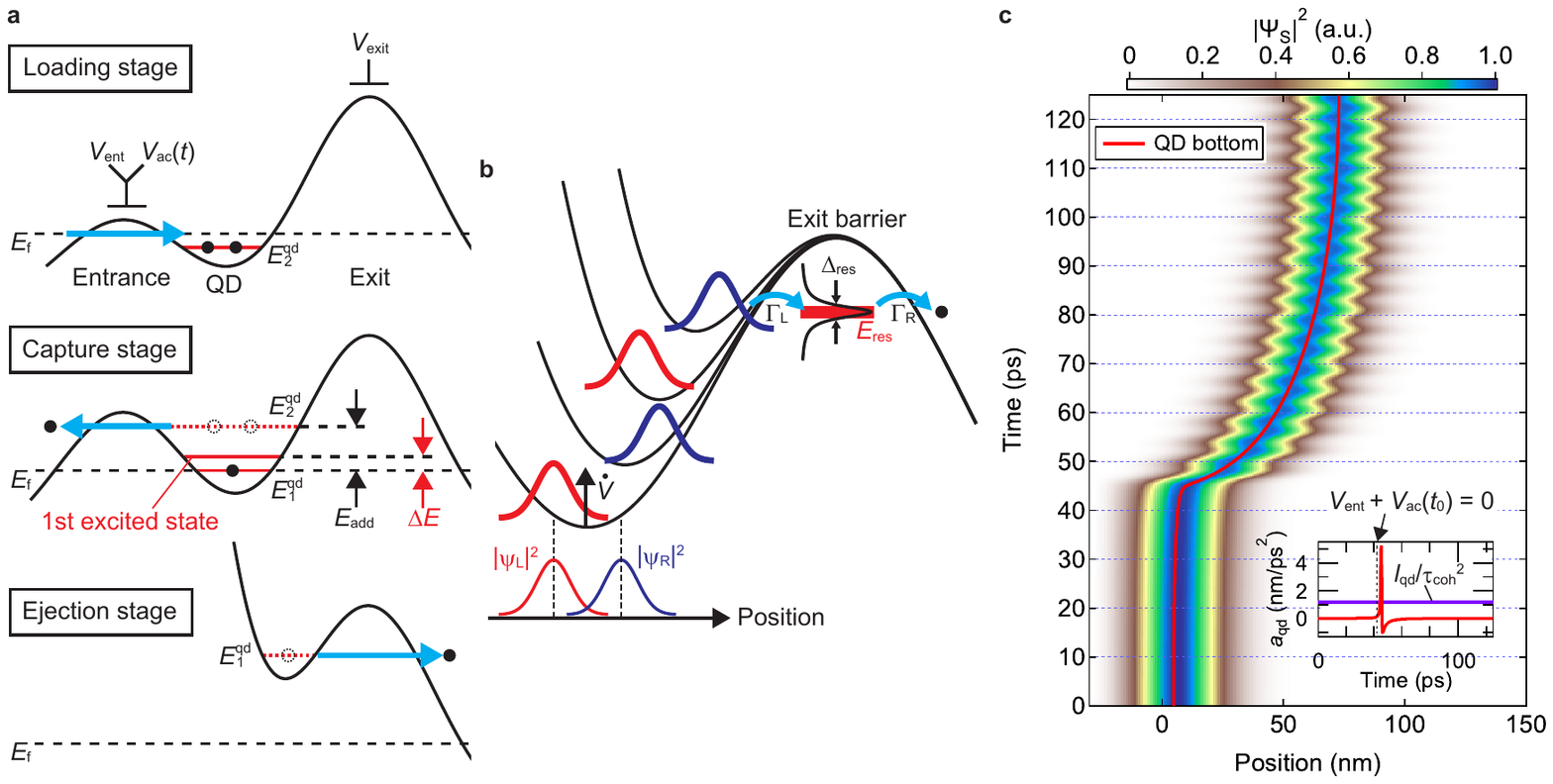}
 \end{center}
 \caption{\bf Internal coherent dynamics in the single-electron source.
 a\rm , Schematic potential diagrams of a quantum dot (QD) electrically formed by applying DC voltages $V_{\mathrm{ent}}$ and $V_{\mathrm{exit}}$. The left and right potential barriers are referred to as the entrance and exit barriers, respectively. $E_{n}^{\mathrm{qd}}$ ($n=1, 2$) is the energy level of the QD with $n$ electrons. The entrance barrier and QD energy level are dynamically tuned by a high-frequency signal $V_{\mathrm{ac}}(t)$, leading to the three stages. $E_{\mathrm{f}}$ and $E_{\mathrm{add}}$, $\Delta E$ are the Fermi level, charge addition energy, and energy gap between the ground and first excited states, respectively.  A single electron is transferred from the left to right leads. \bf b\rm , Schematic diagram of the rise of the QD potential (black solid curves). In the QD, the electron forms a wave packet coherently moving between the left and right sides ($\Ket{\psi_{\mathrm{L}}}=\sqrt{1-p}\Ket{\psi_{\mathrm{G}}} - \sqrt{p}\Ket{\psi_{\mathrm{E}}}$ and $\Ket{\psi_{\mathrm{R}}}=\sqrt{1-p}\Ket{\psi_{\mathrm{G}}} + \sqrt{p}\Ket{\psi_{\mathrm{E}}}$). The spatial distribution $|\psi _{\mathrm{L(R)}}|^2$ of the wave packet is drawn in red (blue) when it is located at the left (right) side. $\dot{V}$ is the rising speed of the QD bottom. The red line in the exit barrier depicts the resonant level $E_{\mathrm{res}}$ with broadening $\Delta _{\mathrm{res}}$. $\Gamma _{\mathrm{L(R)}}$ is the coupling energy between the QD energy level (the right lead) and the resonant level. The wave packet is eventually ejected to the right lead via the resonant level with probability $P_\mathrm{T}$. \bf c\rm, Calculated probability density $|\psi_s|^2$ of the electron wave packet as a function of time and position at $f_{\mathrm{in}}=4$ GHz. 
The inset shows the acceleration of the horizontal movement of the QD as a function of time. The purple line is a critical acceleration for non-adiabatic excitation. The dashed line determines $t_{0}$ for the simulation in Fig. \ref{f5}.}
 \label{f1}
 \end{figure*}

First of all, we explain how coherent oscillations of an electron in a single-electron pump occur. A single-electron pump with a tunable-barrier QD consists of  the entrance (Fig.~\ref{f1}a, left) and exit (right) potential barriers, formed by applying gate voltages $V_{\mathrm{ent}}$ and $V_{\mathrm{exit}}$, respectively\cite{tunable-barrier1}. An AC voltage $V_{\mathrm{ac}}(t)$ with frequency $f_{\mathrm{in}}$ is added to dynamically tune the entrance barrier. The QD energy level is also tuned owing to the cross coupling. When the energy level $E_{n}^{\mathrm{qd}}$ ($n=1,2,\cdots$) of the QD with $n$ electrons is lower than the Fermi level $E_{\mathrm{f}}$, electrons can be loaded from the left lead (loading stage). After that, when $E_{n}^{\mathrm{qd}}$ is lifted and becomes higher than $E_{\mathrm{f}}$, the loaded electrons can escape to the left lead. However, when the escape rate is much slower than the barrier-rise rate, the electrons can be dynamically captured by the QD (capture stage)\cite{kaestner1}. Finally, the captured electrons can be ejected to the right lead (ejection stage). This gives the pumping current as $I_{\mathrm{P}}= nef_{\mathrm{in}}$, where $e$ is the elementary charge. Detailed models for these three stages are found in the Supplementary Information.

When $f_{\mathrm{in}}$ is high (typically in the GHz regime), non-adiabatic excitation can occur in the QD\cite{kataoka1}. Then, electrons can be in a superposition of the ground state and excited states. In the case that only one electron is captured in the QD, the electron forms a wave packet~\cite{kataokaSAW,Sungguen1} moving coherently back and forth between the entrance and exit barriers in the QD, as shown in Fig.~\ref{f1}b. The coherent spatial oscillations can be approximately described by a time-dependent superposition $\Ket{\psi_{\mathrm{S}}(t)}$ between the instantaneous ground state $\Ket{\psi_{\mathrm{G}}(t)}$ and the first excited state $\Ket{\psi_{\mathrm{E}}(t)}$ of the QD,
\begin{equation} \label{SuperpositionGE}
\Ket{\psi_{\mathrm{S}}} \simeq \sqrt{1-p}\Ket{\psi_{\mathrm{G}}}+e^{-i\left ( 2\pi\frac{t}{\tau_{\mathrm{coh}}}-\theta\right)}\sqrt{p}\Ket{\psi_{\mathrm{E}}},
\end{equation}
where $p$ is the probability that the excited state is occupied and $\theta$ is the initial phase. The oscillation period $\tau_{\mathrm{coh}}$ is determined by the energy gap $\Delta E$ between the ground and first excited states, and it is written as $\tau_{\mathrm{coh}}=h/\Delta E$ when $\Delta E$ is approximately time independent.
Here, $h$ is the Planck constant.

To check the feasibility of the non-adiabatic excitation and coherent dynamics described by Eq.~\ref{SuperpositionGE}, we numerically solve the time-dependent Schr$\ddot{\mathrm{o}}$dinger equation with a realistic potential profile (see Supplementary Information). Figure \ref{f1}c shows a calculated wave-packet distribution as a function of time and position at $f_{\mathrm{in}}=4$ GHz. The entrance barrier starts to push the QD away from it at around 40 ps, at which the acceleration $a_{\mathrm{qd}}$ of the horizontal movement of the QD bottom rapidly increases (inset of Fig.~\ref{f1}c). Around this time, the acceleration is higher than the critical value $l_{\mathrm{qd}}/\tau _{\mathrm{coh}}^2$ above which the non-adiabatic excitation occurs, where $l_{\mathrm{qd}}=h/2\pi\sqrt{m\Delta E}$ is the confinement length of the QD and $m$ is the electron effective mass. After that there occurs, following Eq.~\ref{SuperpositionGE}, the spatial oscillations at the picosecond scale, which is beyond currently available bandwidth for measurements of coherent charge oscillations\cite{ntte, Petersson1, DKim1}.

We propose that such fast coherent oscillations can be detected using a resonant level $E_{\mathrm{res}}$ formed in the exit barrier (Fig. \ref{f1}b). While the electron moves back and forth, the potential energy of the QD increases from the value $\overline{E^{\mathrm{qd}}_{\mathrm{ini}}}$ at initial time $t_0$ (onset of the non-adiabatic excitation).
At time $t_1$, when the energy becomes aligned with $E_{\mathrm{res}}$, the electron can be ejected through the exit barrier via the resonant level, generating current.
We formulate the ejection probability $P_{\mathrm{T}}$ based on scattering theory (see Supplementary Information) as
\begin{equation}
P_{\mathrm{T}} \simeq  \overline{P_{\mathrm{T}}} \left [1+2\sqrt{p(1-p)}\mathrm{cos}\left ( 2\pi \frac {t_1 - t_0}{\tau _{\mathrm{coh}}}-\theta \right ) \right ].
\label{coh1}
\end{equation}
$P_{\mathrm{T}}$ depends on the time difference $t_1 - t_0 = (E_{\mathrm{res}}-\overline{E^{\mathrm{qd}}_{\mathrm{ini}}})/\dot{V}$, which is tuned by changing the gate voltages or the rising speed $\dot{V}$ of the QD bottom.
The probability becomes maximal (minimal) when the tuning makes the electron wave packet be located near the exit (entrance) barrier at $t_1$, resulting in gate-dependent current oscillations. 
The mean probability $\overline{P_{\mathrm{T}}} = 2\pi T_{\mathrm{max}} \Delta _{\mathrm{res}} / (\tau_{\mathrm{coh}}\dot{V})$ depends on  the transmission probability $T_{\mathrm{max}}=4\Gamma_{\mathrm{L}}\Gamma_{\mathrm{R}}/(\Gamma_{\mathrm{L}}+\Gamma_{\mathrm{R}})^2$ through the resonant level and on the ratio of the energy broadening $\Delta_{\mathrm{res}}=\Gamma _{\mathrm{L}}+\Gamma _{\mathrm{R}}$ of the resonant level to the energy rise $ \tau_{\mathrm{coh}}\dot{V}$ in one period of the spatial oscillations, where $\Gamma_{\mathrm{L(R)}}$ is the coupling energy between the resonant level and QD (right lead).

The conditions for observing the gate-dependent current oscillations are
\begin{equation}
\Delta  E \lesssim \Delta _{\mathrm{res}} \lesssim \tau _{\mathrm{coh}}\dot{V}.
\label{con1}
\end{equation}
Under the left inequality, the energy uncertainty $\Delta  E$ of the electron wave packet is smaller than the resonance energy broadening $\Delta _{\mathrm{res}}$ so that the electron can fully pass the resonant level\cite{Sungguen1}.
The right inequality is also required; in the opposite limit $\Delta _{\mathrm{res}}/\dot{V} \gg \tau _{\mathrm{coh}}$, the current becomes gate-independent because the wave packet reaches the exit barrier many times within the time window $\Delta _{\mathrm{res}}/\dot{V}$ where the electron is allowed to pass the resonant level (the limit $\Delta _{\mathrm{res}}/\dot{V} \ll \tau _{\mathrm{coh}}$ is also not acceptable because the oscillation amplitude becomes too small, as expected from the expression of $\overline{P_{\mathrm{T}}}$).
  
  \begin{figure*}
\begin{center}
\includegraphics[pagebox=artbox]{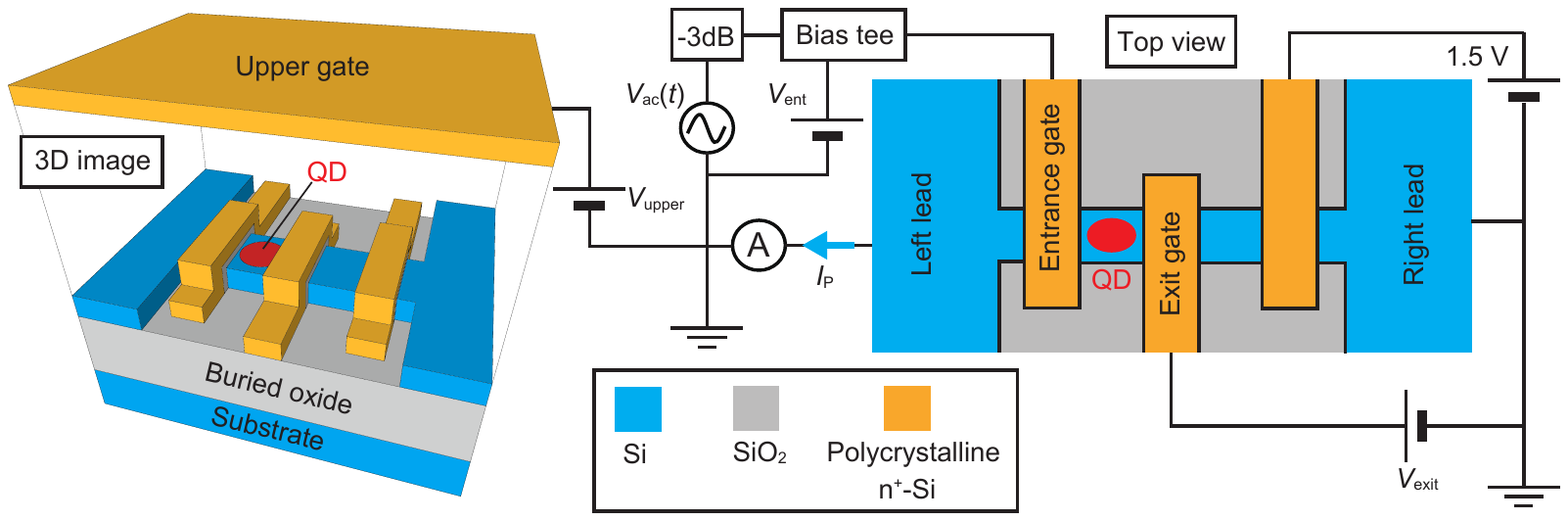}
\end{center}
\caption{\bf Schematic 3D and top images of the device structure. \rm High-frequency signal $V_{\mathrm{ac}}(t)$ with frequency $f_{\mathrm{in}}$ is attenuated by 3 dB and is combined with gate voltage $V_{\mathrm{ent}}$ in the bias tee. The combined signal is applied to the left lower gate (entrance gate). $V_{\mathrm{exit}}$ is applied to the central lower gate (exit gate). 1.5 V is applied to the right lower gate in all experiments in this paper, which turns on the channel under the gate. $V_{\mathrm{upper}}$ is applied to the upper gate. The current through the silicon channel is measured using an ammeter. The red oval indicates the position of the QD.}
\label{f2}
 \end{figure*}
\begin{figure*}
\begin{center}
\includegraphics[pagebox=artbox]{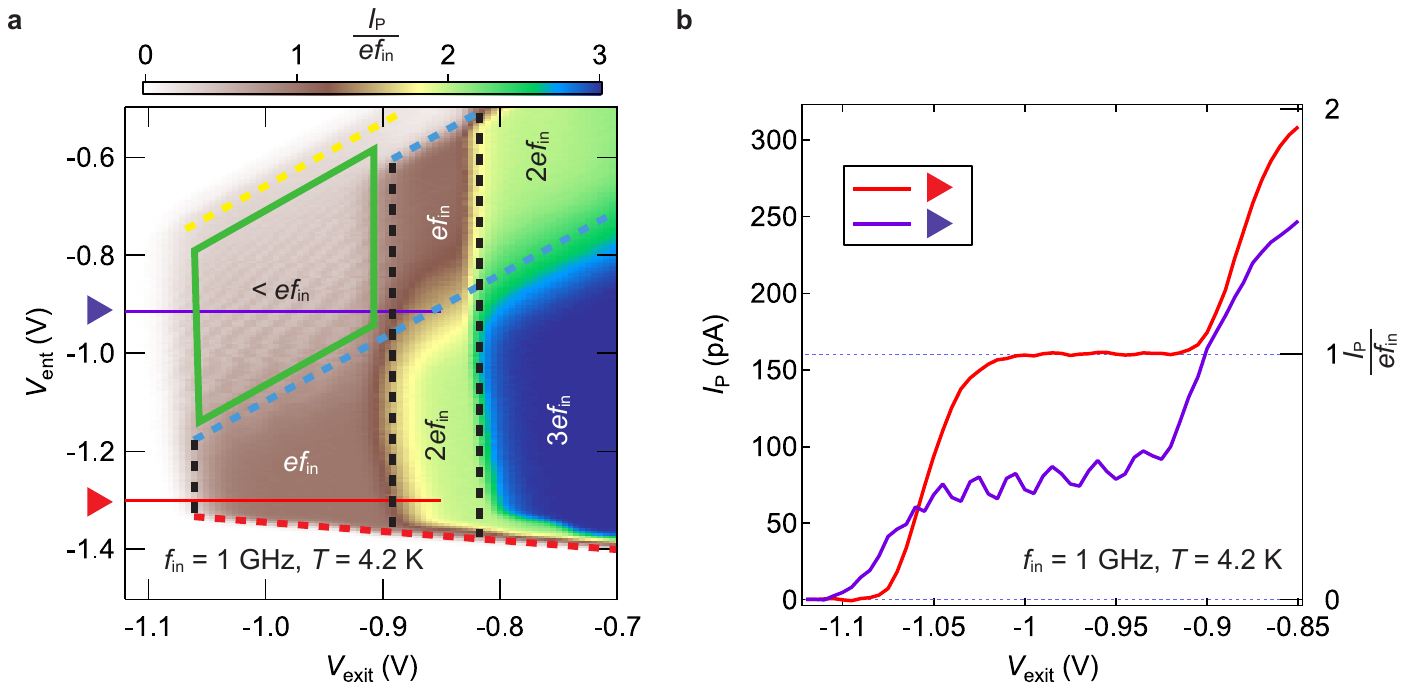}
 \end{center}
 \caption{\bf Experimental observation of current oscillations related to the coherent dynamics. a\rm , Pumping current $I_{\mathrm{P}}$ normalised by $ef_{\mathrm{in}}$ as a function of $V_{\mathrm{ent}}$ and $V_{\mathrm{exit}}$ at $f_{\mathrm{in}}=1$ GHz and $T=4.2$ K, where $V_{\mathrm{upper}}=2.5$ V and power $P$ of the high-frequency signal $V_\mathrm{ac}(t)$ is 10 dBm. The red, black, and blue dashed lines are threshold voltages determined by the loading, capture, and ejection stages, respectively, described in Fig. \ref{f1}a. When the highest QD energy level is lower than $E_{\mathrm{res}}$, the ejection through the resonant level is suppressed, giving another threshold-voltage line (trap-ejection line) indicated by the yellow dashed line. The green parallelogram indicates the region where the current oscillations appear. \bf b\rm ,  $I_{\mathrm{P}}$ (left axis) and $I_{\mathrm{P}}/ef_{\mathrm{in}}$ (right axis) as a function of $V_{\mathrm{exit}}$ at $f_{\mathrm{in}}=1$ GHz and $T=4.2$ K. The red and purple curves are cuts along with the red and purple lines indicated by the red ($V_{\mathrm{ent}}=-1.3$ V) and purple ($V_{\mathrm{ent}}=-0.915$ V) triangles in Fig. \ref{f3}a, respectively.
}
\label{f3}
 \end{figure*}

To observe the coherent dynamics, we measure a device with a double-layer gate structure on a non-doped silicon wire\cite{frat1, NPL-NTT1} at 4.2 K (Fig.~\ref{f2}). The fabrication process and measurement detail are described in Methods. This kind of device often has a resonant level in the exit barrier, which most likely originates from an interface trap level\cite{trap1, trap2, rossi_trap}. Such a resonant level can be identified by investigating a map of $I_{\mathrm{P}}$ as a function of $V_{\mathrm{ent}}$ and $V_{\mathrm{exit}}$ (Fig.~\ref{f3}a). In the map, there are several threshold voltages indicated by dashed lines. Along the red solid line, we observe an $ef_\mathrm{in}$ current plateau (Fig.~\ref{f3}b).
The signature of the resonant level appears in a wide region with a current less than $ef_\mathrm{in}$ indicated by the green parallelogram, where the direct tunneling through the exit barrier is suppressed and the resonant and inelastic tunneling\cite{FujiPh} via the resonant level can be resolved.

\begin{figure*}
\begin{center}
\includegraphics[pagebox=artbox]{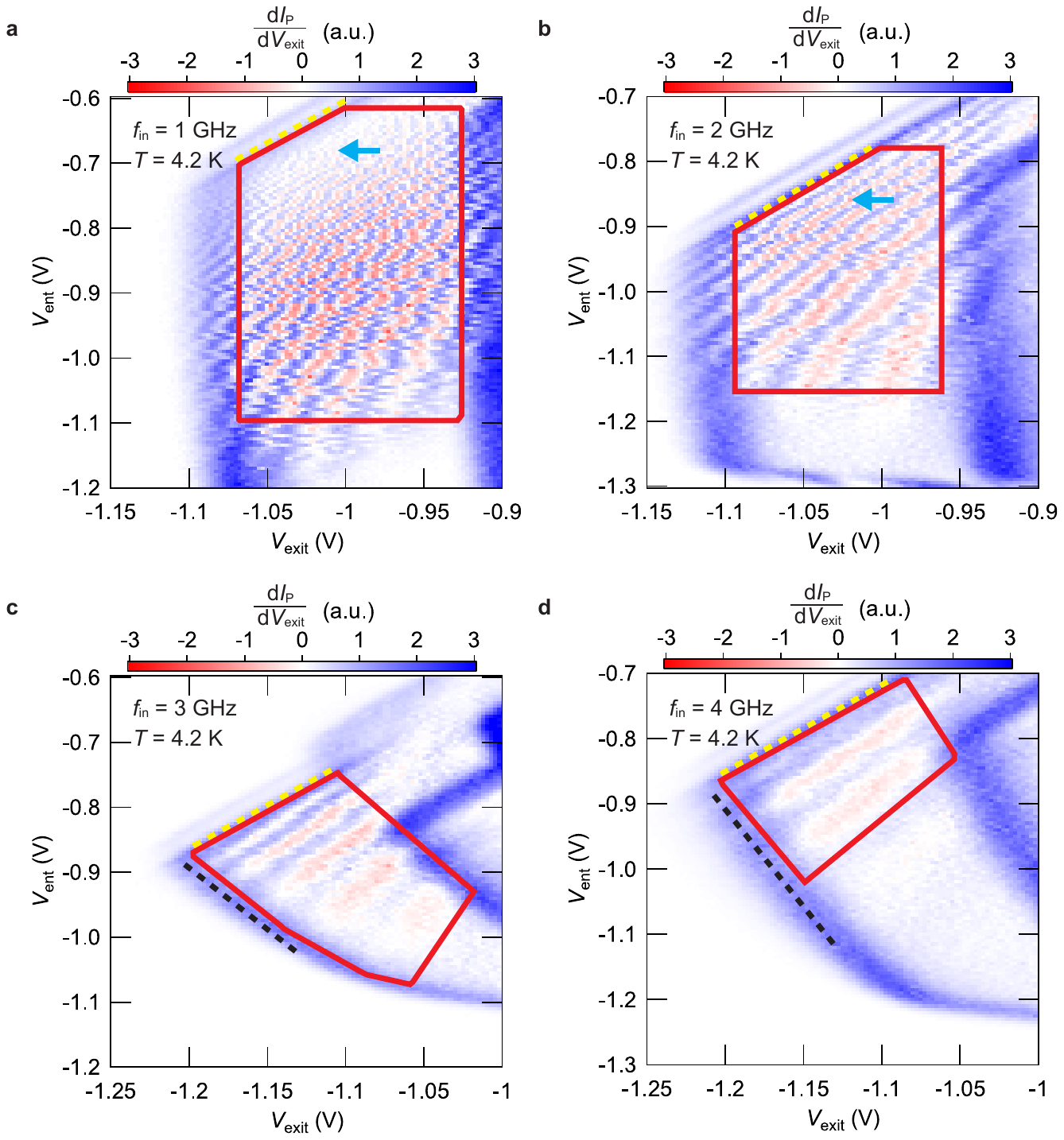}
 \end{center}
 \caption{\bf Frequency dependence of the current oscillations. a-d\rm, First derivative of $I_{\mathrm{P}}$ with respect to $V_{\mathrm{exit}}$ as a function of $V_{\mathrm{ent}}$ and $V_{\mathrm{exit}}$ at $T=4.2$ K and $f_{\mathrm{in}}=1$ GHz (\bf a\rm), 2 GHz (\bf b\rm), 3 GHz (\bf c\rm), and 4 GHz (\bf d\rm), where $V_{\mathrm{upper}}=2.5$ V and $P=10$ dBm. The current oscillations are mainly observed in the region highlighted by red lines. The tilt of the capture line at 3 and 4 GHz indicated by the black dashed lines results from to the cross talk of the high-frequency signal. The trap-ejection lines are indicated by the yellow dashed lines.}
 \label{f4}
 \end{figure*}

We observe current oscillations in the region with the resonant-level signature (see the line cut in Fig.~\ref{f3}b). In this region, the current (normalised by $ef_\mathrm{in}$) through the resonant level increases with increasing $f_{\mathrm{in}}$, indicating the non-adiabatic excitation (the detail is discussed in the Supplementary Information). To examine the current oscillations, we plot $\mathrm{d}I_{\mathrm{P}}/\mathrm{d}V_{\mathrm{exit}}$ as a function of $V_{\mathrm{ent}}$ and $V_{\mathrm{exit}}$ at 1 to 4 GHz (Fig.~\ref{f4}). The period of the current oscillations increases with increasing $f_{\mathrm{in}}$. This is expected from Eq.~\ref{coh1} because $\dot{V}$ increases with increasing $f_{\mathrm{in}}$. In addition, the period becomes shorter when the trap-ejection line (the yellow dashed line) approaches and the oscillation lines slightly bend towards the bottom right. 

\begin{figure*}
\begin{center}
\includegraphics[pagebox=artbox]{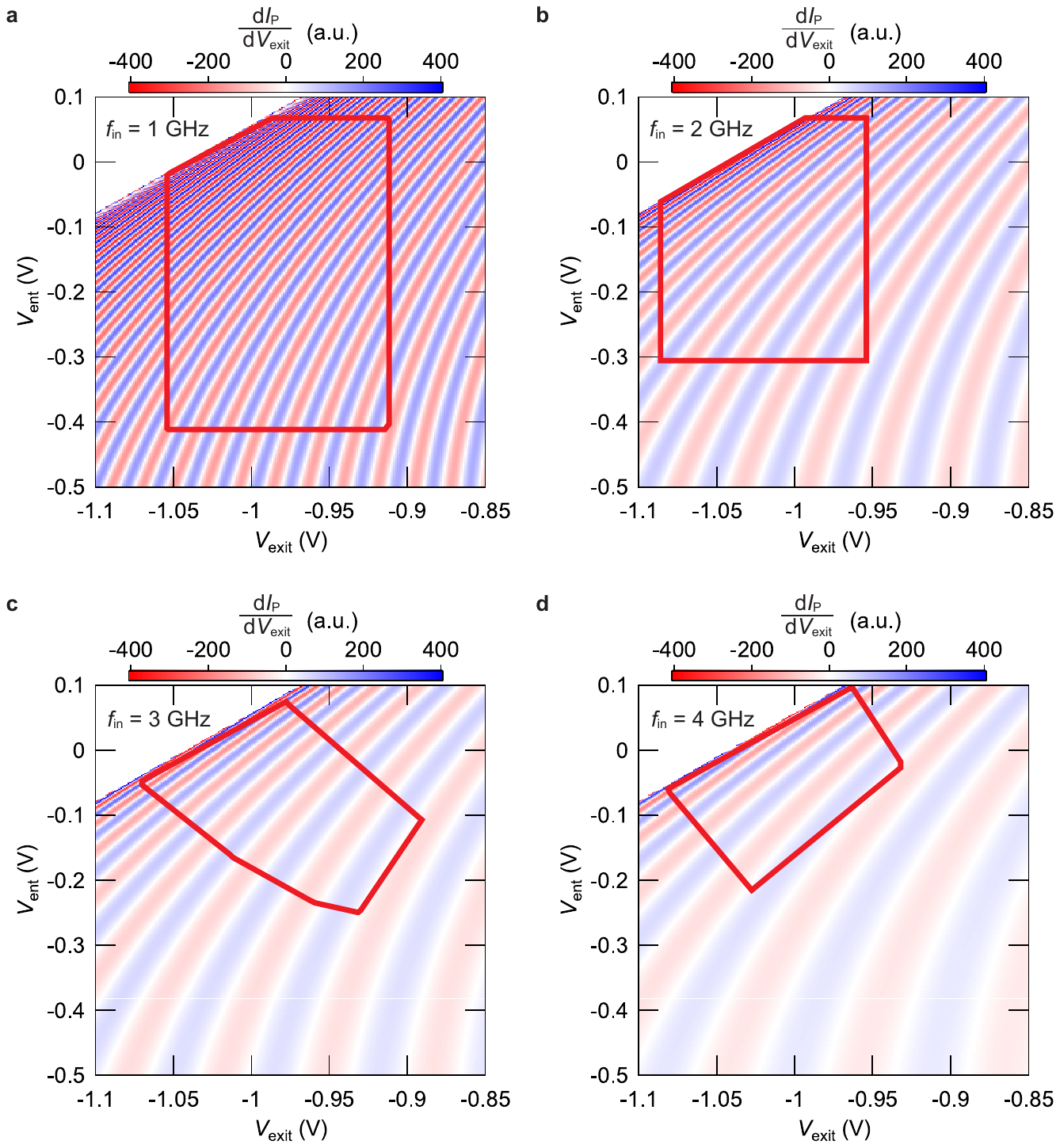}
 \end{center}
 \caption{\bf Calculated frequency dependence of the current oscillations a-d\rm, Calculated first derivative of $I_{\mathrm{P}}$ with respect to $V_{\mathrm{exit}}$ as a function of $V_{\mathrm{ent}}$ and $V_{\mathrm{exit}}$ at $f_{\mathrm{in}}=1$ GHz (\bf a\rm), 2 GHz (\bf b\rm), 3 GHz (\bf c\rm), and 4 GHz (\bf d\rm). The regions highlighted by red lines are the same as those shown in Fig. \ref{f4} at the same $f_{\mathrm{in}}$. $\Delta E=1$ meV in this calculation. The other parameters are estimated from the experimental results (see Supplementary Information).}
 \label{f5}
 \end{figure*}
The above features agree with the coherent spatial oscillations described by Eqs.~\ref{SuperpositionGE} and \ref{coh1}. 
To see this, we simplify Eq.~\ref{coh1} as
\begin{equation}
P_{\mathrm{T}}=\frac{1}{2}\left[ 1 -  \mathrm{cos}\left ( 2\pi \frac {t_{1} - t_{0}}{\tau _{\mathrm{coh}}} \right )\right ],
\end{equation}
which focuses on the position of the current oscillations; the simplification is valid when $p \simeq 0.5$, $\overline{P_{\mathrm{T}}} \simeq $ 0.5, and $\theta \simeq \pi$.
The final time $t_{1}$, at which the energy of the electron becomes aligned with $E_{\mathrm{res}}$, is given by (see Supplementary Information),
\begin{equation}
t_{1}=\frac{1}{2\pi f_{\mathrm{in}}}\mathrm{cos}^{-1}\left[ \frac{1}{V_{\mathrm{amp}}}\left(-V_{\mathrm{ent}}+\frac{\alpha ^{\mathrm{E}}_{\mathrm{exit}}}{\alpha^{\mathrm{E}}_{\mathrm{ent}}}V_{\mathrm{exit}}\right ) \right], 
\end{equation}
where $\alpha ^{\mathrm{E}}_{\mathrm{ent}}$ and $\alpha ^{\mathrm{E}}_{\mathrm{exit}}$ are voltage-to-energy conversion factors related to the entrance and exit gate, respectively. The initial time $t_{0}$ is chosen to be the onset of the non-adiabatic excitation, at which the entrance gate voltage $V_{\mathrm{ent}}+V_{\mathrm{ac}}(t)$ becomes negative so that the entrance barrier starts to push the QD away from it. The onset is determined by $V_{\mathrm{ent}}+V_{\mathrm{ac}}(t_{0})=0$ (see also the inset of Fig.~\ref{f1}c), equivalently
\begin{equation}
t_{0} = \frac{1}{2\pi f_{\mathrm{in}}}\mathrm{cos}^{-1}\left (-\frac{V_{\mathrm{ent}}}{V_{\mathrm{amp}}}\right ).
\end{equation}
By estimating $\alpha ^{\mathrm{E}}_{\mathrm{ent}}$ and $\alpha ^{\mathrm{E}}_{\mathrm{exit}}$ from the measurement results (see Supplementary Information), we simulate the gate dependence of the current oscillations. Figure~\ref{f5} shows calculated current-oscillation maps as a function of $V_{\mathrm{ent}}$ and $V_{\mathrm{exit}}$ at 1 to 4 GHz, corresponding directly to Fig.~\ref{f4}. The peak positions with respect to the trap-ejection line are well reproduced, including the shorter period at the trap-ejection line and the curvature of the oscillation lines (see also discussion in the Supplementary Information).

Importantly, the only fit parameter of the calculation is $\Delta E = 1$ meV. The QD size ($\sim 2l_{\mathrm{qd}}$) estimated from $\Delta E$ is about 40 nm, which is reasonable with respect to the lithographic size of our device (the first electron can be confined at the QD bottom)\cite{nttg1}. The fact that the current oscillations are reproduced using only reasonable parameters supports that the experimental observation is related to the coherent dynamics of the wave packet. We highlight that $\Delta E=1$ meV corresponds to $\tau _{\mathrm{coh}}\sim 4.1$ ps ($1/\tau _{\mathrm{coh}}\sim 240$ GHz), which is far in excess of currently achievable bandwidth\cite{SEsouce1, ntte, Petersson1, DKim1}.

Here, we roughly evaluate Eq.~\ref{con1} using the above results to investigate the validity of the experimental observation (see Supplementary Information), and find that 1 meV $\lesssim \Delta_{\mathrm{res}}\lesssim  6.4$ meV for $\Delta E = 1$ meV and $f_{\mathrm{in}}=1$ GHz. The range of $\Delta_{\mathrm{res}}$ is acceptable, in comparison with the energy difference ($>$ 10 meV, estimated from the result in Fig.~\ref{f3}a) between the top of the exit barrier and the resonant level; the energy difference should be larger than $\Delta_\mathrm{res}$ to observe the current oscillations.

We note that we have considered other possible mechanisms for observed current oscillations, but have not found any alternative explanations. We can neglect the possibilities of the phonon density of states\cite{Phonon_DQD} and the Fabry-Perot interference through an unintentionally-formed QD\cite{FabryPerot1} because they should be independent of $f_{\mathrm{in}}$. We also rule out Landau-Zener-St$\ddot{\mathrm{u}}$ckelberg interference\cite{LZSSi1} (see Supplementary Information).

Our results imply a protocol for measuring such a fast dynamics. We suggest that when any kind of dynamic control of a particle in a cavity, including its initialisation, excitation, and coherent oscillations, is repeated by frequency $f$, the coherent dynamics can be detected as oscillations of the current of the particle through the resonant level, by coupling the cavity to any kind of a resonant level driven by an AC signal with $f$ (see Supplementary Information).

Finally, we stress that the understanding of the internal coherent dynamics is useful for engineering an emitted wave packet, for example, to be a Gaussian form\cite{Sungguen1}. A Gaussian wave packet has a narrow wave-packet width in terms of energy or time (achieving the Heisenberg uncertainty limit), and such a narrow wave packet could lead to ultimately high-speed and high-resolution quantum sensing\cite{Nathan_samp} and enhancement of the visibility of electron quantum optics experiments\cite{Feve_opt1}. Furthermore, since internal coherent dynamics affect the initialisation of flying qubits\cite{yamamoto_fly, MeunSAW1}, this understanding could contribute to the improvement of the initialisation fidelity. In addition, further investigation of the non-adiabatic excitation using the method proposed in this Letter could lead to improvement of the accuracy of single-electron pumping, which contributes to the realisation of high-accuracy current sources.

\section*{Acknowledgements}
We thank K. Chida, H. Tanaka, T. Karasawa, S. P. Giblin, J. D. Fletcher, and T. J. B. M. Janssen for useful discussions. This work was partly supported by JSPS KAKENHI Grant Number JP18H05258 and by the UK Department for Business, Innovation, and Skills and by the EMPIR 15SIB08 e-SI-Amp Project. The latter project has received funding from the EMPIR programme co-financed by the Participating States and from the European Union's Horizon 2020 research and innovation programme. This work was also supported by the National Research Foundation (Korea NRF) funded by Korea Government via the SRC Center for Quantum Coherence in Condensed Matter (Grant Number 2016R1A5A1008184).
 
\section*{Author contributions}
G.Y. measured the device, analyzed the data, and performed the calculation of the pump map. S.R. performed the numerical calculation of the Schr$\ddot{\mathrm{o}}$dinger equation. G.Y. and M.K. conceived the idea of the experiment. S.R. and H.S. developed the theory of the coherent dynamics. N.J. took supporting data of the current oscillations. A.F. fabricated the device. All authors discussed the results. G.Y., S.R., and H.S. wrote the manuscript with feedback from all authors. M.K., A.F., and H.S. supervised the project.

\section*{Competing interests}
The authors declare no competing interests.
\section*{Additional information}
Supplementary information is available for this paper.
\section*{Methods}
\subsection*{Device fabrication}
A silicon wire is patterned using electron beam lithography and dry etching on a non-doped silicon-on-insulator wafer with a buried-oxide thickness of 400 nm. After formation of thermally grown silicon dioxide with a thickness of 30 nm, three lower gates made of heavily-doped polycrystalline silicon are formed using chemical vapor deposition, electron beam lithography, and dry etching. The spacing of the adjacent lower gates is 100 nm. Then, an inter-layer silicon dioxide is deposited using chemical vapor deposition. Next, an upper gate made of heavily-doped polycrystalline silicon is formed using chemical vapor deposition, optical lithography, and dry etching. Then, the left and right leads are heavily doped using ion implantation, during which the upper gate is used as an implantation mask. Finally, aluminum pads are formed to obtain an Ohmic contact. The width and thickness of the silicon wire are 15 and 10 nm, respectively. 
\subsection*{Measurement detail}
Measurements were performed in liquid He at 4.2 K. DC and AC voltages were applied using the Keithley 213 voltage source and Keysight 83623B signal generator, respectively. The pumping current was measured using the Keithley 6514 electrometer. DC voltage $V_{\mathrm{upper}}$ is applied to the upper gate to induce electrons in the silicon wire. $V_{\mathrm{ent}}+V_{\mathrm{ac}}(t)$ and $V_{\mathrm{exit}}$ are applied to the entrance and exit gates, respectively, where $V_{\mathrm{ac}}(t)=V_{\mathrm{amp}}\mathrm{cos}\left (2\pi f_{\mathrm{in}}t\right )$ with amplitude $V_{\mathrm{amp}}$.

\end{document}


\makeatletter 
\renewcommand{\figurename}{Figure}
\renewcommand{\tablename}{Table}
\renewcommand{\refname}{Supporting Reference}
\renewcommand{\thetable}{S\arabic{table}}
\renewcommand{\thefigure}{S\arabic{figure}}
\renewcommand{\bibsection}{}
\renewcommand{\thesubsection}{\Roman{subsection}}
\renewcommand{\theequation}{S\arabic{equation}}
\makeatother

\title{Supplementary information: Picosecond coherent electron motion in a silicon single-electron source} 

%
%
%
%
%

\maketitle
\section{Detailed models of the single-electron pumping}
To estimate the device parameters such as the capacitances from the measurement results, we use detailed models of the single-electron pumping, described below. 

\subsection*{Capture stage: decay cascade model}
The capture of a single electron by the QD during the rise of the QD energy levels  (see Fig. 1a of the main text) can be modeled by a cascade of electron escapes from the QD to left lead: the decay cascade model\cite{kaestner1}. In this model, we can obtain the capture probabilities of electrons by the QD by solving the master equation. When the charge addition energy $E_{\mathrm{add}}$ is large, the capture probabilities $\{$$P_{n-1}^{\mathrm{C}}, P_{n}^{\mathrm{C}}, P_{n+1}^{\mathrm{C}}$$\}$ for $n-1$, $n$, and $n+1$ electrons can be approximated as $\{$$1-e^{-X_{n}^{\mathrm{C}}}, e^{-X_{n}^{\mathrm{C}}}-e^{-X_{n+1}^{\mathrm{C}}} , e^{-X_{n+1}^{\mathrm{C}}}$$\}$, where
\begin{equation}
X_{l}^{\mathrm{C}}=\int_{t^{\mathrm{f}}_{l}}^{t^{\mathrm{E}}} \Gamma^{\mathrm{\mathrm{C}}}_{l}(t)dt \quad \mathrm{for}\quad l=1,2,\cdots,
\label{integC}
\end{equation}
$\Gamma^{\mathrm{C}}_{l}(t)$ is the escape rate of an electron from the QD with $l$ electrons to the left lead, $t^{\mathrm{f}}_{l}$ is the time when the QD energy level with $l$ electrons is aligned with the Fermi level $E_{\mathrm{f}}$ during the rise of the QD energy level, and $t^{\mathrm{E}}$ is the time when the QD energy level is the highest. Using these capture probabilities, we can obtain an expression of the $n$th current plateau determined by the capture as
\begin{equation}
\begin{split}
\frac{I^{\mathrm{C}}}{ef_{\mathrm{in}}}&=(n-1)P_{n-1}^{\mathrm{C}} + nP_{n}^{\mathrm{C}}+(n+1)P_{n+1}^{\mathrm{C}}\\
&= n-1 + e^{-X_{n}^{\mathrm{C}}}+e^{-X_{n+1}^{\mathrm{C}}}.
\label{plateau}
\end{split}
\end{equation}
 
 \begin{figure}
\begin{center}
\includegraphics[pagebox=artbox]{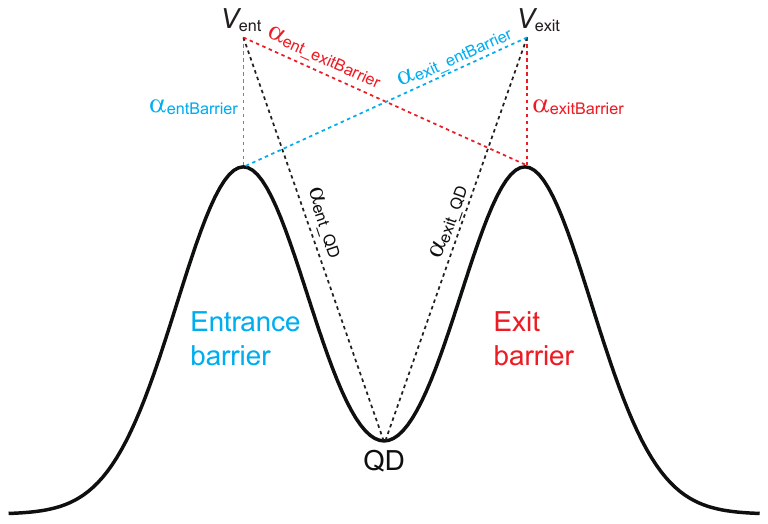}
a\end{center}
 \caption{Definition of the alpha factors, which convert voltage to energy. The gate-QD alpha factors are related to the total QD capacitance $C_{\mathrm{QD}}$: $\alpha _{\mathrm{ent\tiny \_\normalsize QD}}= e\frac{C_{\mathrm{ent\tiny \_\normalsize QD}}}{C_{\mathrm{QD}}}$ and $\alpha _{\mathrm{exit\tiny \_\normalsize QD}}= {e\frac{C_{\mathrm{exit\tiny \_\normalsize QD}}}{C_{\mathrm{QD}}}}$, where $C_{\mathrm{ent(exit)\tiny \_\normalsize QD}}$ is the capacitance between the entrance (exit) gate and QD.}
 \label{f1}
 \end{figure}
To obtain an analytical equation, we simplify the time-dependent voltage applied to the entrance gate as a linear function: $V(t)=-\gamma t$. In this case, the modulation of the entrance barrier, exit barrier, and QD energy level with $l$ electrons can be described as
\begin{eqnarray}
U^{\mathrm{ent}}(t)&=&U_{\mathrm{off}}^{\mathrm{ent}}+\alpha _{\mathrm{entBarrier}}\gamma t,\label{entB}\\
U^{\mathrm{exit}}(t)&=&U_{\mathrm{off}}^{\mathrm{exit}}+ \alpha _{\mathrm{ent\tiny \_\normalsize  exitBarrier}}\gamma t, \label{exitB}\\
E^{\mathrm{qd}}_{l}(t)&=&E^{\mathrm{qd}}_{\mathrm{off}}+\alpha _{\mathrm{ent\tiny \_\normalsize  QD}}\gamma t+(l-1)E_{\mathrm{add}}\label{Edot},
\end{eqnarray}
respectively, where
\begin{eqnarray}
U_{\mathrm{off}}^{\mathrm{ent}}&=&-\alpha _{\mathrm{entBarrier}} V_{\mathrm{ent}}-\alpha _{\mathrm{exit\tiny \_\normalsize entBarrier}} V_{\mathrm{exit}} + U_{0}^{\mathrm{ent}},\label{U1} \\ 
U_{\mathrm{off}}^{\mathrm{exit}}&=&-\alpha _{\mathrm{ent\tiny \_\normalsize  exitBarrier}}V_{\mathrm{ent}}-\alpha _{\mathrm{exitBarrier}}V_{\mathrm{exit}} + U_{0}^{\mathrm{exit}},\label{U2} \\
E^{\mathrm{qd}}_{\mathrm{off}}&=&-\alpha _{\mathrm{ent\tiny \_\normalsize  QD}}V_{\mathrm{ent}}-\alpha _{\mathrm{exit\tiny \_\normalsize QD}} V_{\mathrm{exit}}+E^{\mathrm{qd}}_{0},\label{QDoff}
\end{eqnarray}
$U_{0}^{\mathrm{ent}}$, $U_{0}^{\mathrm{exit}}$, and $E^{\mathrm{qd}}_{0}$ are the constants independent of $V_{\mathrm{ent}}$, $V_{\mathrm{exit}}$, and $l$, and the alpha factors, converting voltage to energy, are depicted in Fig. \ref{f1}. Assuming that an electron tunnels through a parabolic potential barrier\cite{LD}, we can obtain the escape rate as
\begin{equation}
\Gamma^{\mathrm{C}}_{l}(t)=\Gamma^{\mathrm{C}}_{\mathrm{0}} \mathrm{exp}\left [-\frac{U^{\mathrm{ent}}(t)-E^{\mathrm{qd}}_{l}(t)}{k T_{0}}\right ],
\end{equation}
where $\Gamma^{\mathrm{C}}_{\mathrm{0}}$ is the escape rate with a zero entrance-barrier height, $k$ is the Boltzmann constant, and $T_{0}$ is the effective temperature characterized by the tunneling ($T_{0}=\frac{\hbar}{ 2\pi k}\sqrt{\frac{C}{m}}$, where $\hbar$ is the reduced Planck constant, $C$ is the curvature of the barrier, and $m$ is the effective mass of an electron\cite{myPRB1}). Note that $T_{0}$ can be replaced by temperature $T$ at $T > T_{0}$ (thermal hopping). Then, we can calculate Eq. \ref{integC} as
\begin{eqnarray}
X_{l}^{\mathrm{C}}&=&\Gamma^{\mathrm{C}}_{\mathrm{0}}\mathrm{exp}\left [-\frac{U_{\mathrm{off}}^{\mathrm{ent}}-E^{\mathrm{qd}}_{\mathrm{off}}-(l-1) E_{\mathrm{add}}}{kT_{0}}\right ]\int_{t^{\mathrm{f}}_{l}}^{t^{\mathrm{E}}}\mathrm{exp}\left ( - \frac{t}{\tau_{\mathrm{C}}}\right )dt\\
&\sim& \tau_{\mathrm{C}}\Gamma^{\mathrm{C}}_{\mathrm{0}}\mathrm{exp}\left [-\frac{U_{\mathrm{off}}^{\mathrm{ent}}-E^{\mathrm{qd}}_{\mathrm{off}}-(l-1)E_{\mathrm{add}}}{kT_{0}}-\frac{t^{\mathrm{f}}_{l}}{\tau_{\mathrm{C}}}\right ],
\label{xnlinear}
\end{eqnarray}
where
\begin{equation}
\tau_{\mathrm{C}} = \frac{kT_{0}}{\left(\alpha _{\mathrm{entBarrier}} -  \alpha _{\mathrm{ent\tiny \_\normalsize QD}} \right )\gamma}
\end{equation}
and we assume that the escape rate at $t^{\mathrm{f}}_{l}$ is much higher than that at $t^{\mathrm{E}}$. From the condition of $E_{\mathrm{f}}=E^{\mathrm{qd}}_{\mathrm{off}}+\alpha _{\mathrm{ent\tiny \_\normalsize  QD}}\gamma t^{\mathrm{f}}_{l}+ (l-1)E_{\mathrm{add}}$, which means that $t^{\mathrm{f}}_{l}$ depends on the DC gate voltages, we obtain
\begin{equation}
X_{l}^{\mathrm{C}}= \tau_{\mathrm{C}}\Gamma^{\mathrm{C}}_{\mathrm{1}} \mathrm{exp}\left [-\frac{\alpha ^{\mathrm{C}}_{\mathrm{exit}} V_{\mathrm{exit}}- \left (1+1/g\right )(l-1)E_{\mathrm{add}}}{kT_{0}}\right ],
\label{xnlinear2}
\end{equation}
where
\begin{eqnarray}
\alpha ^{\mathrm{C}}_{\mathrm{exit}}&=&\left (1+1/g\right )\alpha _{\mathrm{exit\tiny \_\normalsize QD}}-\alpha _{\mathrm{exit\tiny \_\normalsize entBarrier}},\label{exitQD}\\
g&=&\frac{\alpha _{\mathrm{ent\tiny \_\normalsize QD}}}{\alpha _{\mathrm{entBarrier}}-\alpha _{\mathrm{ent\tiny \_\normalsize QD}}},
\end{eqnarray}
and $\Gamma^{\mathrm{C}}_{\mathrm{1}}$ is the gate-independent constant. Note that $g$ is an important parameter characterising the mechanism of the capture\cite{myPRB1}. Substituting Eq. \ref{xnlinear2} with Eq. \ref{plateau}, we obtain
\begin{equation}
\frac{I^{\mathrm{C}}}{ef_{\mathrm{in}}}=n-1+\sum_{l=n}^{n+1}\mathrm{exp}\left [-\mathrm{exp}\left \{ -\frac{\alpha ^{\mathrm{C}}_{\mathrm{exit}} V_{\mathrm{exit}}-\left (1+1/g\right )(l-1)E_{\mathrm{add}}}{kT_{0}}+\mathrm{ln}(\tau_{\mathrm{C}}\Gamma^{\mathrm{C}}_{\mathrm{1}})\right \} \right ].
\label{cas1}
\end{equation}

\subsection*{Ejection stage}
Ejection of electrons to the right lead can be also modeled using the master equation, which has the same form as that of the capture, with ejection rate $\Gamma^{\mathrm{E}}_{l}(t)$ for $l$ electrons. Therefore, when $n$ electrons are captured in the capture stage, the probabilities with which $l$ electrons are still captured by the QD at the ejection stage are
\begin{eqnarray}
{P}_{n}^{\mathrm{E}}&=&e^{-X_{n}^{\mathrm{E}}},\\
{P}_{l}^{\mathrm{E}}&=&e^{-X_{l}^{\mathrm{E}}}-e^{-X_{l+1}^{\mathrm{E}}} \quad \mathrm{for}\quad  1\leq l<n,\\
{P}_{0}^{\mathrm{E}}&=&1-e^{-X_{1}^{\mathrm{E}}},
\end{eqnarray}
where
\begin{equation}
X_{l}^{\mathrm{E}}=\int_{t^{\mathrm{f}}_{l}}^{t^{\mathrm{E}}} \Gamma^{\mathrm{\mathrm{E}}}_{l}(t)dt.
\label{ejein}
\end{equation}
In this case, the current determined by the ejection is
\begin{equation}
\frac{I^{\mathrm{E}}}{ef_{\mathrm{in}}} =  \sum_{l=0}^{n} (n-l) P_{l}^{\mathrm{E}}  = n-\sum_{l=1}^{n}e^{-X_{l}^{\mathrm{E}}}.
\label{ejeI}
\end{equation}

Similar to the capture stage, the ejection rate can be obtained using Eqs. \ref{exitB} and \ref{Edot} as
\begin{equation}
\Gamma^{\mathrm{E}}_{l}(t)=\Gamma^{\mathrm{E}}_{\mathrm{0}} \mathrm{exp}\left [-\frac{U^{\mathrm{exit}}(t)-E^{\mathrm{qd}}_{l}(t)}{kT_{0}}\right ],
\end{equation}
where $\Gamma^{\mathrm{E}}_{\mathrm{0}}$ is the ejection rate with a zero exit-barrier height. Then, we calculate Eq. \ref{ejein} as
\begin{eqnarray}
X_{l}^{\mathrm{E}}&=&\Gamma^{\mathrm{E}}_{\mathrm{0}}\mathrm{exp}\left [-\frac{U_{\mathrm{off}}^{\mathrm{exit}}-E^{\mathrm{qd}}_{\mathrm{off}}-(l-1) E_{\mathrm{add}}}{kT_{0}}\right ]\int_{t^{\mathrm{f}}_{l}}^{t^{\mathrm{E}}}\mathrm{exp}\left (\frac{t}{\tau_{\mathrm{E}}}\right )dt\\
&\sim& \tau_{\mathrm{E}}\Gamma^{\mathrm{E}}_{\mathrm{0}}\mathrm{exp}\left [-\frac{U_{\mathrm{off}}^{\mathrm{exit}}-E^{\mathrm{qd}}_{\mathrm{off}}-(l-1) E_{\mathrm{add}}}{kT_{0}}+\frac{t^{\mathrm{E}}}{\tau_{\mathrm{E}}}\right ],
\end{eqnarray}
where
\begin{equation}
\tau_{\mathrm{E}} = \frac{kT_{0}}{\left(\alpha _{\mathrm{ent\tiny \_\normalsize QD}} -  \alpha _{\mathrm{ent\tiny \_\normalsize exitBarrier}} \right )\gamma}
\end{equation}
and we assume that the ejection rate at $t^{\mathrm{E}}$ is much higher than that at $t^{\mathrm{f}}_{l}$. Since $t^{\mathrm{E}}$ is independent of the DC gate voltages, we obtain
\begin{equation}
X^{\mathrm{\mathrm{E}}}_{l}= \tau_{\mathrm{E}} \Gamma^{\mathrm{\mathrm{E}}}_{\mathrm{1}}\mathrm{exp}\left [-\frac{\alpha ^{\mathrm{E}}_{\mathrm{ent}}V_{\mathrm{ent}} - \alpha ^{\mathrm{E}}_{\mathrm{exit}}V_{\mathrm{exit}}- (l-1)E_{\mathrm{add}}}{kT_{0}}  \right ],
\label{ejeGam}
\end{equation}
where
\begin{eqnarray}
\alpha ^{\mathrm{E}}_{\mathrm{ent}}&=&\alpha _{\mathrm{ent\tiny \_\normalsize QD}}-\alpha _{\mathrm{ent\tiny \_\normalsize exitBarrier}},\label{ejeEnt}\\
\alpha ^{\mathrm{E}}_{\mathrm{exit}}&=&\alpha _{\mathrm{exitBarrier}}-\alpha _{\mathrm{exit\tiny \_\normalsize QD}},\label{ejeExit}
\end{eqnarray}
and $\Gamma^{\mathrm{E}}_{\mathrm{1}}$ is the gate-independent constant. 
Substituting Eq. \ref{ejeGam} with Eq. \ref{ejeI}, we obtain
\begin{equation}
\frac{I^{\mathrm{E}}}{ef_{\mathrm{in}}} = n-\sum_{l=1}^{n}\mathrm{exp}\left [ -\mathrm{exp}\left \{-\frac{\alpha ^{\mathrm{E}}_{\mathrm{ent}} V_{\mathrm{ent}} - \alpha ^{\mathrm{E}}_{\mathrm{exit}}V_{\mathrm{exit}}- (l-1) E_{\mathrm{add}}}{kT_{0}}+\mathrm{ln}(\tau_{\mathrm{E}}\Gamma^{\mathrm{E}}_{\mathrm{1}})\right \}   \right ].
\label{ejeI1}
\end{equation}

\subsection*{Loading stage}
The probability of the initial loading of electrons at the lowest QD energy level can be determined by the alignment between the QD energy level and Fermi level. In this case, we should observe multiple loading lines reflecting $E_{\mathrm{add}}$. However, this is not the case in our device because our results show that the loading line is shared with all plateaus, which is usually observed in other devices\cite{kataoka1, NPL-NTT1}. Thus, it would be determined by the loading rate through the entrance barrier. Since the lowest QD energy level should be much deeper than the Fermi level, we assume for simplicity that the loading from the left lead with the loading rate $\Gamma^{\mathrm{L}}(t)$ only contributes. For the evaluation of the alpha factors related to the entrance barrier, it would be enough to consider the master equation of the loading probability ${P}^{\mathrm{L}}_{1}$ of the first electron, which can be written as
\begin{equation}
\frac{\mathrm{d}{P}^{\mathrm{L}}_{1}}{\mathrm{d}t}=\Gamma^{\mathrm{L}}(t)\left ( 1-{P}_{1}^{\mathrm{L}} \right).
\end{equation}
The solution of this equation is
\begin{equation}
{P}^{\mathrm{L}}_{1}=1-\mathrm{exp}\left( -\int_{\tilde{t}^{\mathrm{f}}_{1}}^{t^{\mathrm{L}}} \Gamma^{\mathrm{\mathrm{L}}}(t)dt \right ),
\label{xlinteg}
\end{equation}
where $\tilde{t}^{\mathrm{f}}_{1}$ is the time when the QD energy level for the first electron is aligned with the Fermi level during the fall of the QD energy level and $t^{\mathrm{L}}$ is the time when the QD energy level is lowest. 

Since the fall of the QD energy level contributes in this case, we assume the linear time-dependent voltage as $V(t)=\gamma t$, which changes Eq. \ref{entB} to
\begin{equation}
\tilde{U}^{\mathrm{ent}}(t)=U_{\mathrm{off}}^{\mathrm{ent}}-\alpha _{\mathrm{entBarrier}}\gamma t.
\label{entB2}
\end{equation}
Similar to the above two stages, we obtain the loading rate using Eq. \ref{entB2} as
\begin{equation}
\Gamma^{\mathrm{L}}(t)=\Gamma^{\mathrm{L}}_{\mathrm{0}}\mathrm{exp}\left (-\frac{\tilde{U}^{\mathrm{ent}}(t)-E_{\mathrm{f}}}{kT_{0}}\right ),
\end{equation}
where $\Gamma^{\mathrm{L}}_{0}$ is the loading rate with a zero entrance-barrier height. Then we calculate the integral of Eq. \ref{xlinteg} as
\begin{eqnarray}
\int_{\tilde{t}^{\mathrm{f}}_{1}}^{t^{\mathrm{L}}} \Gamma^{\mathrm{\mathrm{L}}}(t)dt&=& \Gamma^{\mathrm{L}}_{\mathrm{0}}\mathrm{exp}\left (-\frac{U_{\mathrm{off}}^{\mathrm{ent}}-E_{\mathrm{f}}}{kT_{0}}\right )\int_{\tilde{t}^{\mathrm{f}}_{1}}^{t^{\mathrm{L}}}\mathrm{exp}\left (\frac{t}{\tau_{\mathrm{L}}}\right )dt\\
&\sim& \tau_{\mathrm{L}}\Gamma^{\mathrm{L}}_{\mathrm{0}}\mathrm{exp}\left [-\frac{U_{\mathrm{off}}^{\mathrm{ent}}-E_{\mathrm{f}}}{kT_{0}}+\frac{t^{\mathrm{L}}}{\tau_{\mathrm{L}}}\right ],
\end{eqnarray}
where
\begin{equation}
\tau_{\mathrm{L}} = \frac{kT_{0}}{\alpha _{\mathrm{entBarrier}} \gamma}
\end{equation}
and we assume that the loading rate at $t^{\mathrm{L}}$ is much higher than that at $\tilde{t}^{\mathrm{f}}_{1}$. Since $t^{\mathrm{L}}$ is independent of the DC gate voltages, we obtain
\begin{equation}
\int_{\tilde{t}^{\mathrm{f}}_{1}}^{t^{\mathrm{L}}} \Gamma^{\mathrm{\mathrm{L}}}(t)dt= \tau_{\mathrm{L}} \Gamma^{\mathrm{L}}_{1}\mathrm{exp}\left (\frac{\alpha _{\mathrm{entBarrier}} V_{\mathrm{ent}} + \alpha _{\mathrm{exit\tiny \_\normalsize entBarrier}} V_{\mathrm{exit}}}{kT_{0}}  \right ),
\label{integLoad}
\end{equation}
where $\Gamma^{\mathrm{L}}_{1}$ is the gate-independent constant. Then, we obtain the current determined by the loading as
\begin{equation}
\frac{I^{\mathrm{L}}}{ef_{\mathrm{in}}} ={P}^{\mathrm{L}}_{1}
= 1-\mathrm{exp}\left [ -\mathrm{exp}\left \{\frac{\alpha _{\mathrm{entBarrier}}V_{\mathrm{ent}} + \alpha _{\mathrm{exit\tiny \_\normalsize entBarrier}}  V_{\mathrm{exit}}}{kT_{0}}+\mathrm{ln}(\tau_{\mathrm{L}} \Gamma^{\mathrm{L}}_{\mathrm{1}})\right \}   \right ].
\label{loadL}
\end{equation}

\section{Estimation of device parameters}
\subsection*{Charge addition energy $\boldsymbol{E_{\mathrm{add}}}$}
\begin{figure}[h]
\begin{center}
\includegraphics[pagebox=artbox]{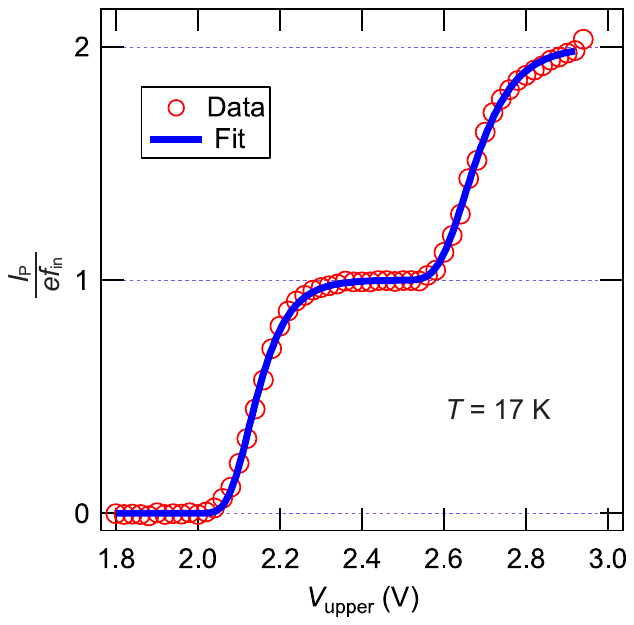}
\end{center}
\caption{$I_{\mathrm{P}}$ normalised by $ef_{\mathrm{in}}$ as a function of $V_{\mathrm{upper}}$ at $f_{\mathrm{in}}=50$ MHz and $T=17$ K, where $V_{\mathrm{exit}}=-1$ V. We use voltage pulses as a high-frequency signal, where high and low voltages are 0 and $-3$ V, respectively. The fit curve yields $\alpha ^{\mathrm{C}}_{\mathrm{upper}1}=0.028$ eV/V, $\alpha ^{\mathrm{C}}_{\mathrm{upper}2}=0.022$ eV/V, $V_{1}=2.1$ V, and $V_{2}=2.7$ V. }
\label{f2}
 \end{figure}
$E_{\mathrm{add}}$ is equal to the charging energy $E_{\mathrm{C}}=e^{2}/C_{\mathrm{QD}}$ at the $ef_{\mathrm{in}}$ plateau, where $C_{\mathrm{QD}}$ is the total capacitance of the QD, because of the spin degeneracy. To estimate $E_{\mathrm{C}}$, we use the upper gate dependence of the current plateau at a high temperature of 17 K (Fig. \ref{f2}). For evaluation of the $ef_{\mathrm{in}}$ plateau, we change the alpha factor of the exit gate to that of the upper gate in Eq. \ref{cas1} and simplify the equation as
\begin{equation}
\frac{I^{\mathrm{C}}}{ef_{\mathrm{in}}}=\sum_{l=1}^{2} \mathrm{exp}\left [-\mathrm{exp}\left \{ -\frac{\alpha ^{\mathrm{C}}_{\mathrm{upper}}\left( V_{\mathrm{upper}}-V_{l}\right )}{kT}\right \} \right ],
\label{casfit}
\end{equation}
where $V_{l}$ is the threshold voltage of the $l$th plateau. In this case, $\alpha ^{\mathrm{C}}_{\mathrm{upper}}(V_{2}-V_{1})=(1+1/g)E_{\mathrm{C}}$. However, the experimental results have different alpha factors for the first and second plateaus possibly because of the gate dependence of the capacitances. Thus, we use different alpha factors ($\alpha ^{\mathrm{C}}_{\mathrm{upper}1}$ and $\alpha ^{\mathrm{C}}_{\mathrm{upper}2}$ for the first and second plateaus, respectively) and the averaged value of them for the estimation of $E_{\mathrm{C}}$ instead of $\alpha ^{\mathrm{C}}_{\mathrm{upper}}$. Another difficulty is the $g$ value, which can not be directly estimated from Fig. \ref{f2}. The $g$ value is an indicator of the pumping mechanism of the capture: the decay cascade model is suitable at $g \gg 1$ and the thermal equilibrium model is suitable at $g \ll1$\cite{myPRB1}. Here, we assume that the mechanism is close to the decay cascade model, which is typical in our device\cite{frat1}, and $g$ is assumed to be 10 (see Fig. 5f in Ref.~3). Note that the contribution of $g$ is small at $g>10$ because the factor used in the estimation is $1+1/g$. From the parameters extracting the fitting, we obtain $E_{\mathrm{C}}=12$ meV and $C_{\mathrm{QD}}=14$ aF.

\subsection*{Barrier modulation}
\begin{figure}[h]
\begin{center}
\includegraphics[pagebox=artbox]{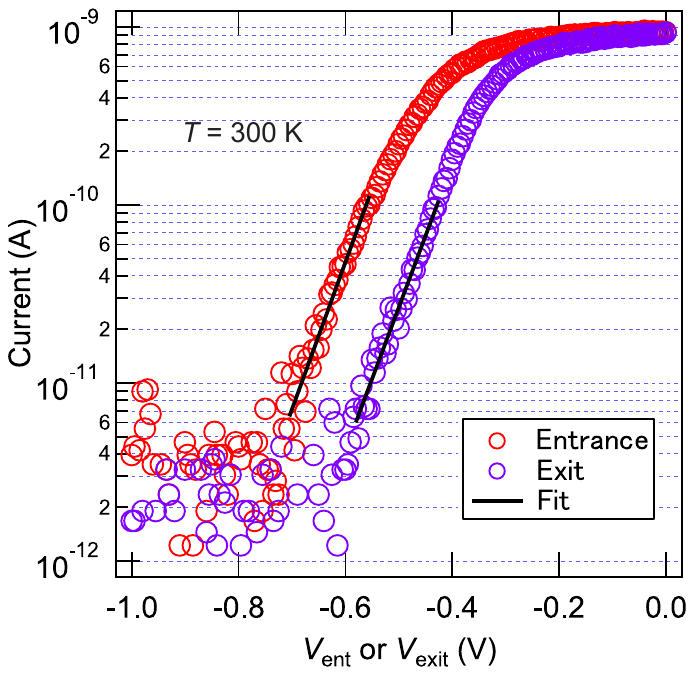}
 \end{center}
 \caption{DC current as a function of $V_{\mathrm{ent}}$ (red circles) and $V_{\mathrm{exit}}$ (purple circles) at $T=300$ K, where voltages applied to the other gates are 1 V. The DC bias is almost zero (only an offset voltage of the ammeter). From the fit lines, the subthreshold slopes are estimated, resulting in $1.2\times 10^{2}$ mV/decade for the entrance ($S_{\mathrm{ent}}$) and exit ($S_{\mathrm{exit}}$) gates.}
\label{f3}
 \end{figure}

Since the modulations of the entrance and exit barriers by the entrance and exit gates, respectively, are just the operation of the transistors, the alpha factors related to them can be estimated from the subthreshold slope $S_{\mathrm{ent (exit)}}$ of the transistors: $\alpha _{\mathrm{entBarrier(exitBarrier)}}=\frac{kT\mathrm{ln}(10)}{S_{\mathrm{ent (exit)}}}$. Figure \ref{f3} shows DC characteristics of the entrance and exit gates in the subthreshold regime at 300 K, where the black lines are linear fits of $\mathrm{ln}(\mathrm{Current})$. From the parameters extracted from the fits, $\alpha _{\mathrm{entBarrier}}= 0.49$ eV/V and $\alpha _{\mathrm{exitBarrier}}= 0.48$ eV/V.

\subsection*{Cross couplings for entrance \& exit barriers}
\begin{figure}[h]
\begin{center}
\includegraphics[pagebox=artbox]{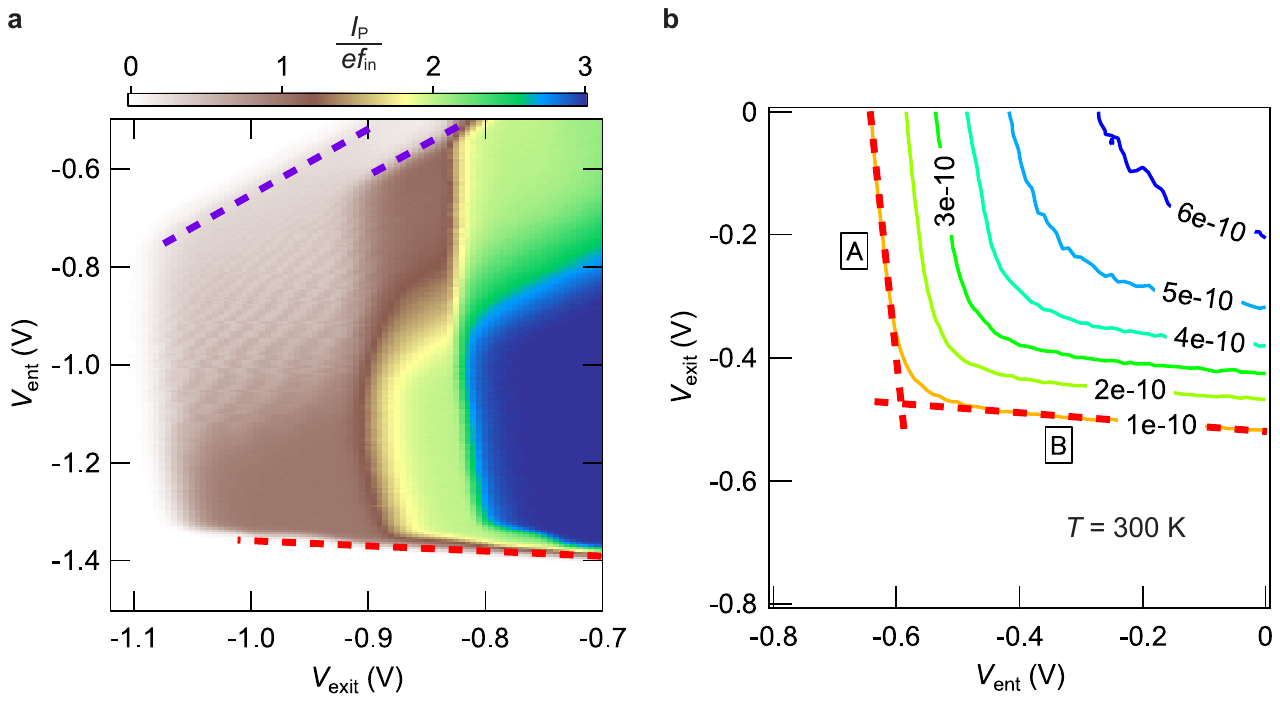}
 \end{center}
 \caption{\bf a\rm, The same current map as shown in Fig. 3a in the main text. The slopes of the loading (red dashed line) and (trap-)ejection (purple dashed line) lines are $-0.11$ and $1.3$, respectively. \bf b\rm, Contour plot of a DC current of a device with the same structure as a function of $V_{\mathrm{exit}}$ and $V_{\mathrm{ent}}$ at $T=300$ K, where $V_{\mathrm{upper}}=2$ V, a voltage applied to the right lower gate (see Fig. 2 in the main text) is 1 V, the DC bias is 10 mV. The slopes of A and B indicated by the red dashed lines are -9.4 and -0.076, respectively.}
 \label{f4}
 \end{figure}

$\alpha _{\mathrm{exit\tiny \_\normalsize entBarrier}}$ can be estimated from the loading lines because Eq. \ref{loadL} contains $\alpha _{\mathrm{exit\tiny \_\normalsize entBarrier}}$ and $\alpha _{\mathrm{entBarrier}}$. The red dashed line in Fig. \ref{f4}a is the loading lines on the $V_{\mathrm{ent}}$-$V_{\mathrm{exit}}$ map. $\alpha _{\mathrm{exit\tiny \_\normalsize entBarrier}}$ is the product of $\alpha _{\mathrm{entBarrier}}$ and the absolute value of the slope of the loading line. From the estimated values of the slopes, $\alpha _{\mathrm{exit\tiny \_\normalsize entBarrier}}=0.052$ eV/V.

$\alpha _{\mathrm{ent\tiny \_\normalsize exitBarrier}}$ can be estimated from DC transport characteristics but we use a different way to extract it because of the lack of the data. Here, we use the data of another device that was fabricated on the same wafer with the same design as the device in this paper. Figure \ref{f4}b shows a contour plot of a DC current as a function of $V_{\mathrm{ent}}$ and $V_{\mathrm{exit}}$ at 300 K. From the slope A, we estimate $\alpha _{\mathrm{exit\tiny \_\normalsize entBarrier}}/\alpha _{\mathrm{entBarrier}}=0.11$, which is the same value as the absolute value of the slope of the loading line shown in Fig. \ref{f4}a, indicating the same device structure. Then, from the slope B, $\alpha _{\mathrm{ent\tiny \_\normalsize exitBarrier}}/\alpha _{\mathrm{exitBarrier}}=0.076$, leading to  $\alpha _{\mathrm{ent\tiny \_\normalsize exitBarrier}}=0.037$ eV/V. Note that the slight asymmetric value of the slopes A and B would results from the small difference of the gate length.

\subsection*{Gate-QD couplings}
 \begin{figure}[b]
\begin{center}
\includegraphics[pagebox=artbox]{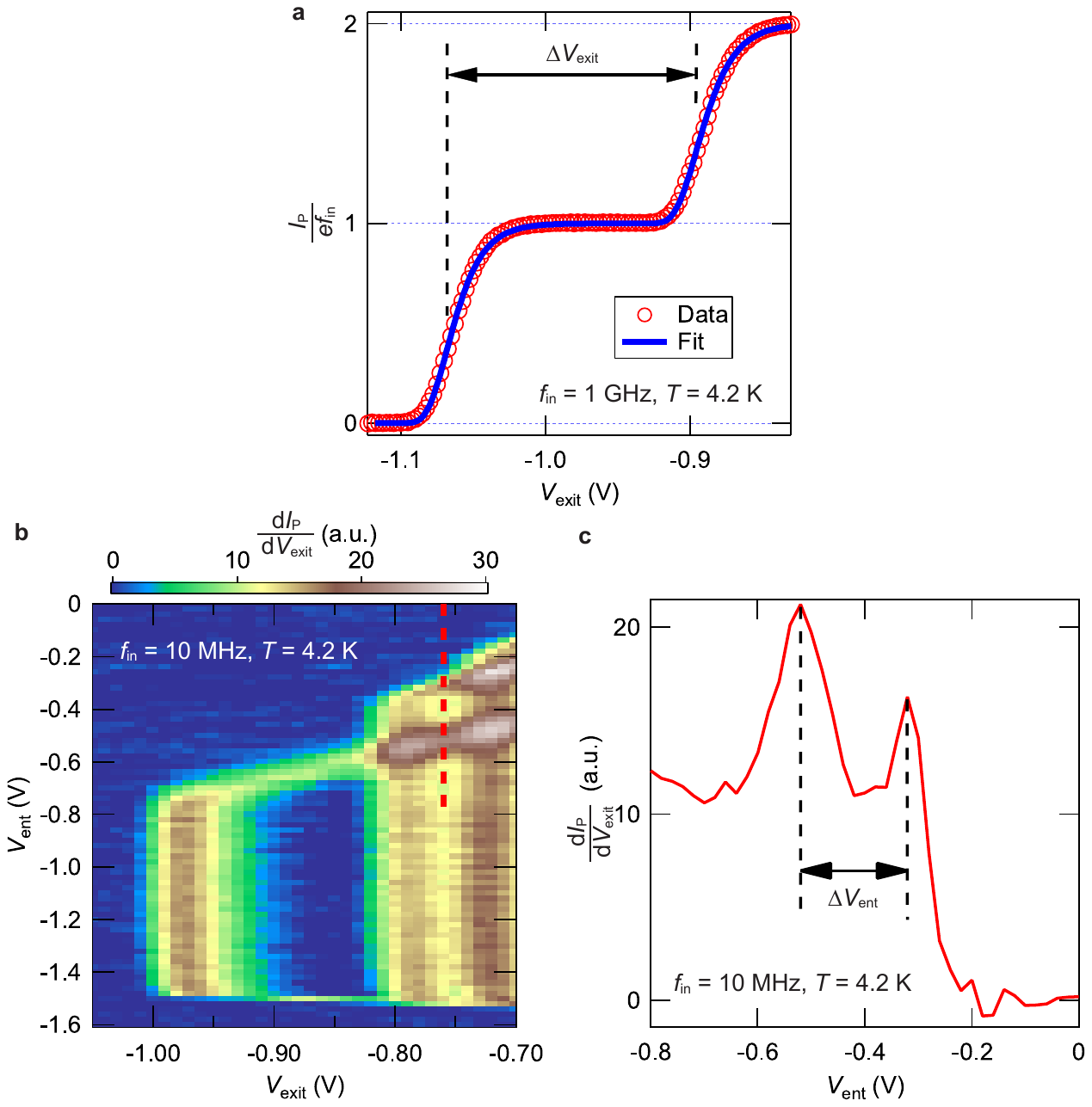}
 \end{center}
 \caption{\bf a\rm, $I_{\mathrm{P}}$ normalised by $ef_{\mathrm{in}}$ as a function of $V_{\mathrm{exit}}$ at $f_{\mathrm{in}}=1$ GHz and $T=4.2$ K, where $V_{\mathrm{ent}}=-1.25$ V, $V_{\mathrm{upper}}=2.5$ V, and $P=10$ dBm. The fit curve yields $\Delta V_{\mathrm{exit}}$ of  $1.7$ V. \bf b\rm, First derivative of $I_{\mathrm{P}}$ with respective to $V_{\mathrm{exit}}$ as a function of $V_{\mathrm{ent}}$ and $V_{\mathrm{exit}}$ at $f_{\mathrm{in}}=10$ MHz and $T=4.2$ K, where $V_{\mathrm{upper}}=2.5$ V and $P=9$ dBm. \bf c\rm, Cut along with the red dashed line in Fig. \ref{f5}b, where $V_{\mathrm{exit}}=-0.76$ V. $\Delta V_{\mathrm{ent}}$ is estimated to be $0.20$ V.}
 \label{f5}
 \end{figure}
The gate-QD couplings at the capture stage ($\alpha _{\mathrm{ent\tiny \_\normalsize QD}}^{\mathrm{C}}$, $\alpha _{\mathrm{exit\tiny \_\normalsize QD}}^{\mathrm{C}}$) should be different from those at the ejection stage ($\alpha _{\mathrm{ent\tiny \_\normalsize QD}}^{\mathrm{E}}$, $\alpha _{\mathrm{exit\tiny \_\normalsize QD}}^{\mathrm{E}}$) because the potential profile is largely different. Thus, we separately extract them. 

From the assumed $g$ value of 10, $\alpha _{\mathrm{ent\tiny \_\normalsize QD}}^{\mathrm{C}}=\alpha _{\mathrm{entBarrier}}/(1+1/g)=0.45$ eV/V. Figure \ref{f5}a shows a normalized current as a function of $V_{\mathrm{exit}}$ at 1 GHz, where the spacing between the threshold voltages of the first and second plateaus is $\Delta V_{\mathrm{exit}}$ extracted from the fit using
\begin{equation}
\frac{I^{\mathrm{C}}}{ef_{\mathrm{in}}}=\sum_{l=1}^{2} \mathrm{exp}\left [-\mathrm{exp}\left \{ -\frac{\alpha ^{\mathrm{C}}_{\mathrm{exit}}\left( V_{\mathrm{exit}}-V_{l}\right )}{kT_{0}}\right \} \right ],
\label{casfit2}
\end{equation}
which is similar to Eq. \ref{casfit}. From a relation of $[ (1+1/g)\alpha _{\mathrm{exit\tiny \_\normalsize QD}}^{\mathrm{C}}-\alpha _{\mathrm{exit\tiny \_\normalsize entBarrier}}]\Delta V_{\mathrm{exit}}=(1+1/g)E_{\mathrm{add}}$ (see Eqs. \ref{exitQD}, \ref{cas1}), $\alpha _{\mathrm{exit\tiny \_\normalsize QD}}^{\mathrm{C}}=0.10$ eV/V. The capacitances between the gates and QD in the capture stage are as follows: $C_{\mathrm{ent\tiny \_\normalsize QD}}^{\mathrm{C}}= C_{\mathrm{QD}}\alpha _{\mathrm{ent\tiny \_\normalsize QD}}^{\mathrm{C}}/e = 6.0$ aF and $C_{\mathrm{exit\tiny \_\normalsize QD}}^{\mathrm{C}}= C_{\mathrm{QD}}\alpha _{\mathrm{exit\tiny \_\normalsize QD}}^{\mathrm{C}}/e = 1.4$ aF, for the entrance and exit gates, respectively.

Since the ejection line is not clear at a high frequency because of the inelastic current through the resonant level discussed below, we use the 10-MHz data, which has clear ejection lines, to extract the spacing between the ejection lines of the first and second electrons $\Delta V_{\mathrm{ent}}$ (Figs. \ref{f5}b and \ref{f5}c). From the relation of $(\alpha _{\mathrm{ent\tiny \_\normalsize QD}}^{\mathrm{E}}-\alpha _{\mathrm{ent\tiny \_\normalsize exitBarrier}})\Delta V_{\mathrm{ent}}=E_{\mathrm{add}}$ (see Eqs. \ref{ejeEnt} and \ref{ejeI1}), we obtain $\alpha _{\mathrm{ent\tiny \_\normalsize QD}}^{\mathrm{E}}=0.096$ eV/V. The slope indicated by the purple dashed lines in Fig. \ref{f4}a is equal to $(\alpha _{\mathrm{exitBarrier}}-\alpha _{\mathrm{exit\tiny \_\normalsize QD}}^{\mathrm{E}})/(\alpha _{\mathrm{ent\tiny \_\normalsize QD}}^{\mathrm{E}}-\alpha _{\mathrm{ent\tiny \_\normalsize exitBarrier}})$ (see Eqs. \ref{ejeEnt}, \ref{ejeExit}, and \ref{ejeI1}), leading to $\alpha _{\mathrm{exit\tiny \_\normalsize QD}}^{\mathrm{E}}=0.40$ eV/V. The capacitances between the gates and QD in the ejection stage are as follows: $C_{\mathrm{ent\tiny \_\normalsize QD}}^{\mathrm{E}}= C_{\mathrm{QD}} \alpha _{\mathrm{ent\tiny \_\normalsize QD}}^{\mathrm{E}}/e = 1.3$ aF and $C_{\mathrm{exit\tiny \_\normalsize QD}}^{\mathrm{E}}= C_{\mathrm{QD}}\alpha _{\mathrm{exit\tiny \_\normalsize QD}}^{\mathrm{E}}/e= 5.4$ aF for the entrance and exit gates, respectively.

\subsection*{Discussion}
The alpha factors, capacitances, and $E_{\mathrm{add}}$ are summarised in Table \ref{tab:price}. The change in the gate-QD alpha factors from the capture stage to the ejection one indicates that the QD moves from the entrance to exit barriers. The fact that $\alpha _{\mathrm{ent\tiny \_\normalsize QD}}^{\mathrm{C}}\sim \alpha _{\mathrm{exit\tiny \_\normalsize QD}}^{\mathrm{E}}$ indicates that the relative position between the entrance gate and QD at the capture stage is similar to that between the exit gate and QD at the ejection stage. From $\alpha _{\mathrm{exitBarrier}}$ and  $\alpha _{\mathrm{exit\tiny \_\normalsize QD}}^{\mathrm{E}}$, the $g$ value of the exit gate can be estimated to be about 5. These facts would indicate that the assumption of $g=10$ for the entrance gate is not so bad. 

\begin{table}[htb]
  \begin{center}
    \caption{Summary of the alpha factors, addition energy, and capacitances}
    \begin{tabular}{||c|c||c|c|c|c||} \hline
      \multicolumn{2}{||c||}{Alpha factors} & \multicolumn{2}{c|}{Alpha factors  \& $E_{\mathrm{add}}$} &  \multicolumn{2}{|c||}{Capacitances} \\ \hline
      $\alpha _{\mathrm{entBarrier}}$&   0.49 eV/V   & $\alpha _{\mathrm{ent\tiny \_\normalsize QD}}^{\mathrm{C}}$ & 0.45 eV/V& $C_{\mathrm{ent\tiny \_\normalsize QD}}^{\mathrm{C}}$ & 6.0 aF\\
      $\alpha _{\mathrm{exitBarrier}}$&  0.48 eV/V  & $\alpha _{\mathrm{exit\tiny \_\normalsize QD}}^{\mathrm{C}}$ & 0.10 eV/V& $C_{\mathrm{exit\tiny \_\normalsize QD}}^{\mathrm{C}}$ & 1.4 aF\\
      $\alpha _{\mathrm{exit\tiny \_\normalsize entBarrier}}$ & 0.052 eV/V  &$\alpha _{\mathrm{ent\tiny \_\normalsize QD}}^{\mathrm{E}}$  &0.096 eV/V&$C_{\mathrm{ent\tiny \_\normalsize QD}}^{\mathrm{E}}$  & 1.3 aF \\
      $\alpha _{\mathrm{ent\tiny \_\normalsize exitBarrier}}$ & 0.037 eV/V&$\alpha _{\mathrm{exit\tiny \_\normalsize QD}}^{\mathrm{E}}$  & 0.40 eV/V &$C_{\mathrm{exit\tiny \_\normalsize QD}}^{\mathrm{E}}$ & 5.4 aF\\
      -&-& $E_{\mathrm{add}}$ &12 meV&$C_{\mathrm{QD}}$&14 aF\\ \hline
    \end{tabular}
    \label{tab:price}
  \end{center}
\end{table}

\section{Numerical calculation of time-dependent schr\mbox{\boldmath $\ddot{\mathrm{O}}$}dinger equation}

We here show that the nonadiabatic excitation and coherent oscillations, described by Eq. 1 of the main text, can be induced in the experimental setup in Fig. 2 of the main text.  We first show that the adiabatic condition is violated in the experiment, based on rough estimation from the experimental parameters. Then we numerically calculate the Schr\"{o}dinger equation governed by a realistic potential profile of the time-dependent QD. The result (Fig.~1c in the main text) shows the coherent oscillations of Eq. 1.  In addition, calculations with respect to various gate voltages confirm that the onset of the non-adiabatic excitation is determined by Eq.~6 of the main text.

\begin{figure}[h]
\begin{center}
\includegraphics[width=0.85\textwidth]{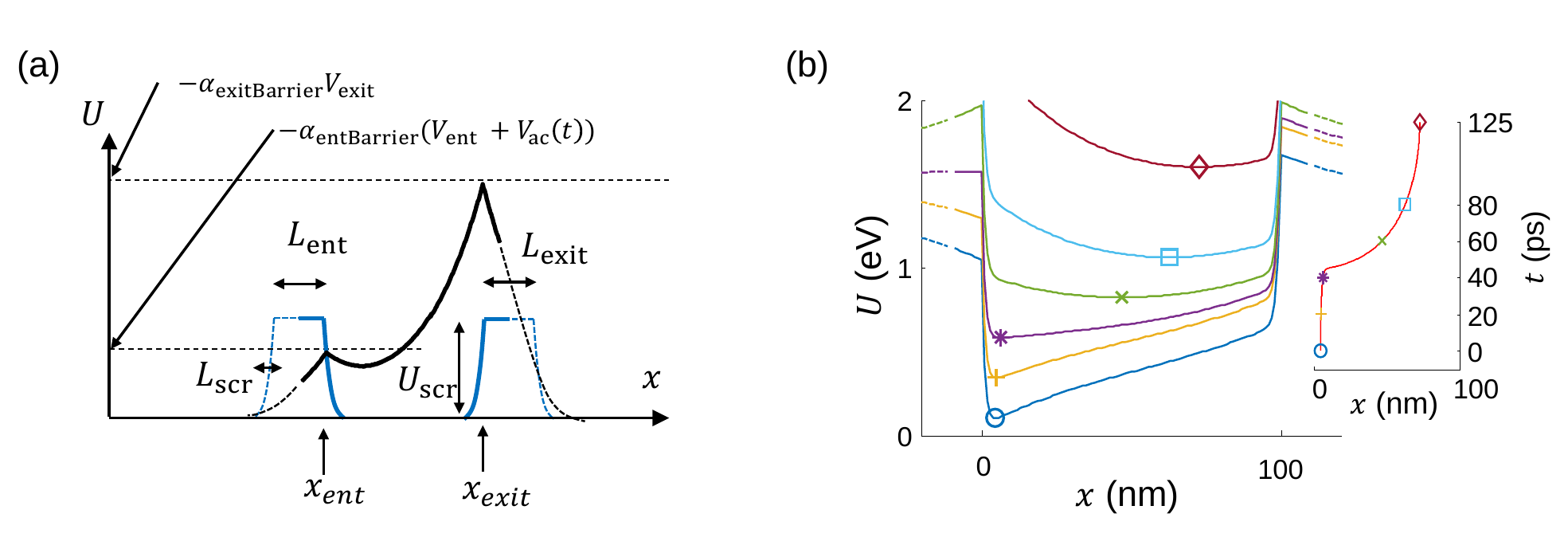}
 \end{center}
 \caption{\bf a\rm, \bf b\rm, Schematic diagram (\bf a\rm ) and the potential profile (\bf b\rm ) described by Eq.~\ref{eq:pot_profile}. In (\bf b\rm ), the potential is drawn at times 0, 20, 40, 60, 80, 125 ps from the bottom to the top, with choosing the realistic parameters of $f=4$ GHz, $V_\text{ent}=-0.7$ V, $V_\text{exit}= -0.7$ V, $V_\text{amp} = 1.415$ V,  $x_\text{ent}=0$, and $x_\text{exit}=100 $ nm.  The plots are vertically shifted for clarity. In the inset, the position of the QD potential minimum at the selected times (see the markers) is shown.  }
 \label{fig:supp_QD_potential}
 \end{figure}

We show that the adiabatic condition is violated in the experiment, based on rough estimation from the experimental parameters. For the purpose, we focus on the spatial movement of the QD (see Fig.~1b in the main text). 
The adiabatic condition is $\delta x/\delta t \ll l_\text{qd} /\tau_\text{coh}$, which means that the QD moves distance $\delta x$, much shorter than $l_\text{qd}$, during the time $\delta t$ much longer than $h/\Delta E$. We consider a simple QD potential $U(x,t)$ based on the experimentally estimated parameters. The simplification is that the potential is parabolic near the potential minimum and the potential induced by the entrance gate linearly decreases from the entrance barrier side to the potential minimum, $U(x,t) = \frac{\Delta E}{2 l_\mathrm{qd}^2}  (x-x_0)^2 + (\alpha_\mathrm{ent\_QD}-\alpha_\mathrm{entBarrier})[V_\mathrm{ac}(t)-V_\mathrm{ac}(t_0)] x/x_0$. Here, $x$ is measured from the entrance barrier, $x_0$ is the distance between the QD potential minimum and the entrance barrier at time $t_0$. According to the potential, the velocity $\delta x/ \delta t$ is estimated as
$\delta x/\delta t =  ( \alpha_\text{entBarrier} - \alpha_\mathrm{ent\_QD} ) l_\text{qd}^2 |\dot{V}_\text{ac}| /(x_0 \Delta E) $,  where $\dot{V}_\text{ac} = d V_\text{ac} / dt$.
With the realistic parameters, we find $\delta x/\delta t \simeq  $ 9.7 nm/ps, which is about two times larger than $l_\text{qd}/\tau_\text{coh} =  $ 4.9 nm/ps. Here we used the parameters of  $\alpha_\text{ent\_QD}^\text{C} $, $\alpha_\text{ent\_QD}^\text{E} $ and $\alpha_\text{entBarrier}$ shown in Table ~\ref{tab:price}, $\alpha_\mathrm{ent\_QD} = (\alpha_\text{ent\_QD}^\text{C}+\alpha_\text{ent\_QD}^\text{E})/2 $ (which is chosen as rough average during the pumping), $ |\dot{V}_\text{ac}| =  $ 5.6  mV/ps (which is the average value of  $\dot{V}_\text{ac}$ during the first half of the pumping with $f=$ 1 GHz and $V_\text{amp} = 1.4$ V), $x_{0} = $ 50 nm roughly obtained from the gate geometry, $\Delta E = $ 1 meV as estimated in the main manuscript, and $l_\text{qd} = $ 20 nm is determined by $\Delta E$ and the electron effective mass $m = 0.19 m_e$ with the bare mass $m_e$. The estimation implies that nonadiabatic excitation occurs but not too strongly so that the occupation of the first excited state of the QD dominates over those of the other excited states (see Eq.~1 of the main text). Note that this estimation is based on the essential parameters characterizing the pump, rather than relying on the details of the potential profile.

Next, we explain the potential profile of the dynamic QD that is used in the numerical calculations of the time-dependent Schr\"{o}dinger equation. The potential profile is contributed from the potentials induced by the entrance, exit, and upper gates [see Fig.~\ref{fig:supp_QD_potential}(a)],
\begin{equation}
  \label{eq:pot_profile}
  \begin{aligned}
  U(x,t) &= U_\text{ent} (x,t) + U_\text{exit} (x) + U_\text{upper} (x)  \\
  U_\text{ent}(x,t) &= -\alpha_\text{entBarrier} \left[V_\text{ent}+V_\text{ac}(t)\right]  \left(\frac{\alpha_\text{entBarrier}}{\alpha_\text{ent\_exitBarrier}} \right)^{-\frac{ |x-x_\text{ent}|}{|x_\text{exit}-x_\text{ent}|} } \\
  U_\text{exit}(x) &= -\alpha_\text{exitBarrier} V_\text{exit} \left( \frac{\alpha_\text{exitBarrier}}{\alpha_\text{exit\_entBarrier}} \right)^{- \frac{ |x-x_\text{exit}|}{|x_\text{ent}-x_\text{exit}|} } \\
  U_\text{upper}(x) &= U_\text{scr} \exp \left[ -\frac{x -x_\text{ent}}{ L_\text{scr} } \Theta (x-x_\text{ent}) \right] \exp \left[  -\frac{x_\text{ent}-L_\text{ent}-x }{ L_\text{scr} } \Theta (x_\text{ent} -L_\text{ent} -x) \right] \\
  & \qquad +  U_\text{scr} \exp \left[ -\frac{x -x_\text{exit}-L_\text{exit}}{ L_\text{scr} } \Theta (x-x_\text{ent}-L_\text{exit}) \right] \exp \left[  -\frac{x_\text{exit}-x }{ L_\text{scr} } \Theta (x_\text{exit}  -x) \right]
  \end{aligned}
\end{equation}
The potentials $U_\text{ent}$ induced by the entrance gate and $U_\text{exit}$ by the exit gate together form the QD in the region between the gate edge positions $x_\text{ent}$ and $x_\text{exit}$ [see the black curve in Fig.~\ref{fig:supp_QD_potential}(a)]. Away from the gate edges, the potential exponentially decays with the decay length determined by the alpha factors in Table~\ref{tab:price}. The profile in Eq.~\ref{eq:pot_profile} captures this feature: $U_\text{ent}$ has the value of $-\alpha_\text{entBarrier} (V_\text{ent}+V_\text{ac}(t))$ at $x = x_\text{ent}$ and exponentially decays to $-\alpha_\text{ent\_exitBarrier} (V_\text{ent}+V_\text{ac}(t)) $ at $x= x_\text{exit}$. Similarly, $U_\text{exit}$ has the value of $-\alpha_\text{exitBarrier} V_\text{exit}$ at $x= x_\text{exit}$ and exponentially decays to $- \alpha_\text{exit\_entBarrier} V_\text{exit}$ at $x= x_\text{ent}$. On the other hand, the upper gate induces the potential $U_\text{upper}$ nontrivially, since the entrance and exit gates screen the electric field induced by the upper gate (see Fig.~2 of the main text). For positive $V_\text{upper}$, the screening reduces the effect of the upper gate in the region underneath the entrance (exit) gates, resulting in the formation of the two potential barrier contributions corresponding to the two terms of $U_\text{upper}(x)$ in Eq.~\ref{eq:pot_profile}. The barriers have width $L_\text{ent}$ and $L_\text{exit}$, respectively, and exponentially decay from the maximum value $U_\text{scr}$ within length $L_\text{scr}$ [see the blue curve in Fig.~\ref{fig:supp_QD_potential}(a)]. 
We roughly choose the values of $U_\text{scr}=$ 1 eV and $L_\text{scr}=$ 1 nm.
We note that the time evolution determined by the potential profile is well described by Eqs. 1 and 6 of the main text, insensitively to the detailed values of  $U_\text{scr}$ and $L_\text{scr}$. In addition, the detailed shape of
the outer parts of the potential profile [shown as the black dashed curves and the blue dashed curves in Fig.~\ref{fig:supp_QD_potential}(a)] does not affect the coherent oscillations of the wave packet when the non-adiabatic excitation occurs not too strongly.
 
The potential profile of Eq.~\ref{eq:pot_profile} evolves in time as in Fig.~\ref{fig:supp_QD_potential}(b) during a half period of pumping cycle.  During the time from 0 to 40 ps [see the lowest three curves in Fig.~\ref{fig:supp_QD_potential}(b)], the QD is formed at $x_\text{ent}$ near the entrance gate. 
At around 40 ps, the QD starts to be pushed away from $x_\text{ent}$ because of the formation of the entrance barrier. 
The potential bottom of the QD follows the trajectory shown as the red curve in the inset of Fig.~\ref{fig:supp_QD_potential}(b). 
This spatial movement of the QD is the dominant factor resulting in the nonadiabatic excitation; we observe that the excitation probability becomes $\sim 10^{-2}$ times smaller if the spatial shift of the QD bottom is artificially compensated by a shift of the whole potential profile.

We compute the time evolution of the wave packet inside the QD, shown in Fig.~1c, solving the time-dependent Schr\"{o}dinger equation. The spatial movement of the QD, the origin of the nonadiabatic excitation, is taken into account as follows: We simplify the QD potential such that the potential profile has time-independent parabolic shape but its potential minimum moves following the trajectory shown in the inset of Fig.~\ref{fig:supp_QD_potential}(b); the trajectory is determined by the realistic parameters and Eq.~\ref{eq:pot_profile} as discussed above. 
The level spacing of the QD is chosen as $\Delta E = $ 1 meV. The initial wave function at $t=0$ is chosen as the ground state wave packet of the potential profile at the initial time $t=0$. 
We note that the calculation result of the time evolution is insensitive to the choice of the initial time as long as the acceleration of the QD at the initial time is much slower than $l_\text{qd}/\tau_\text{coh}^2$. 

The time evolution shown in Fig.~1c is obtained at 4 GHz pumping frequency.  The nonadiabatic excitation occurs around the time of 40 ps, as the acceleration of the QD movement becomes faster than $l_\text{qd}/\tau_\text{coh}^2$; see the inset of Fig.~1c. After that, the time evolution is well described by Eq.~1 in the main text.
The contribution of the higher excited states of the QD to the nonadiabatic evolution is $\sim 10^{-2}$ smaller than that of the first excited state.

\begin{figure}[h]
\begin{center}
\includegraphics[width=0.85\textwidth]{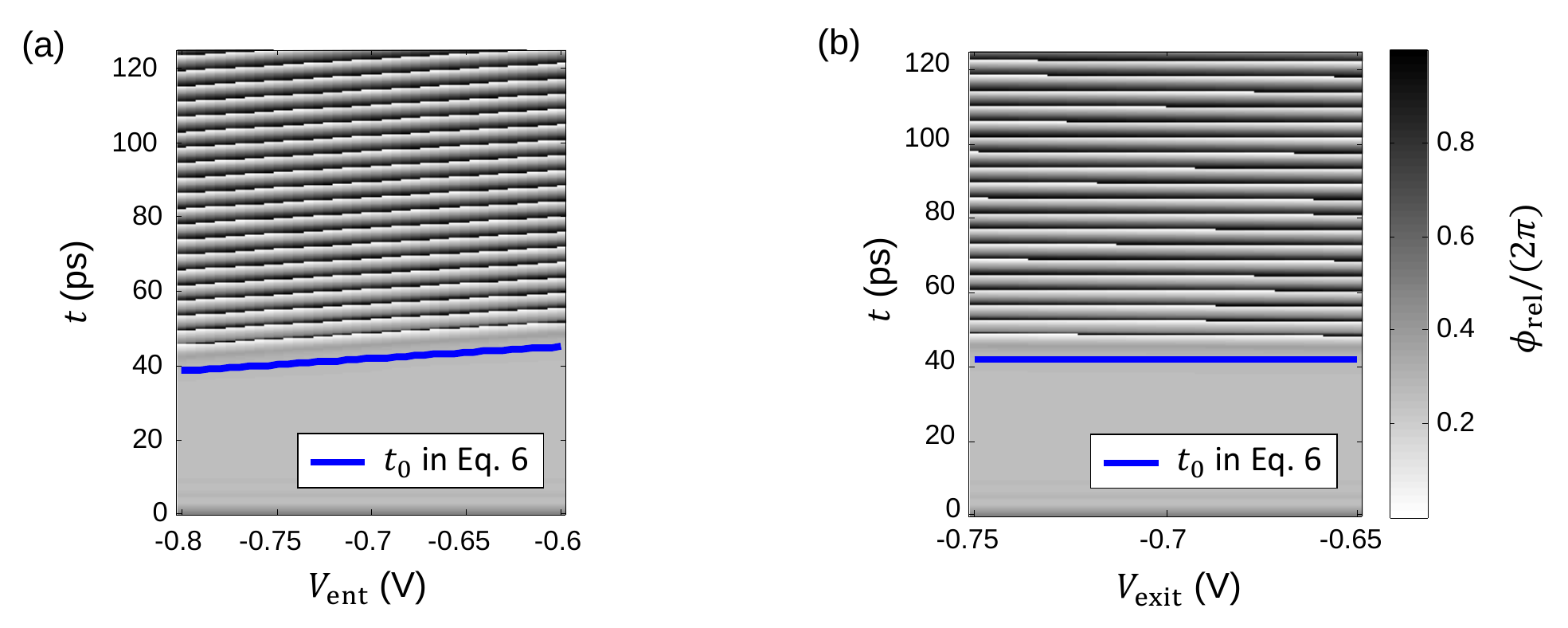}
 \end{center}
 \caption{\bf a\rm, \bf b\rm, The relative phase $\phi_\text{rel}$ of the coherent oscillations for different gate voltages. We choose $V_\text{exit} = -0.7$ V in (\bf a\rm) and $V_\text{ent}=-0.7$ V in (\bf b\rm). The phase is shown in the range of $ [0, 2\pi]$. The blue line is the onset of the nonadiabatic excitation; see Eq.~6 in the main text.}
 \label{fig:supp_phase}
 \end{figure}

We calculate the time evolution of the wave packet for different values of  $V_\text{ent}$ and $V_\text{exit}$ in Fig.~\ref{fig:supp_phase}, which shows the relative phase  $\phi_\text{rel}(t) \equiv  \text{angle} [\langle \psi_\text{E} | \psi_S (t)\rangle / \langle \psi_\text{G} | \psi_S (t)\rangle]$ between the ground state and the first excited state in the superposition representing the time evolution (see Eq.~1 in the main text). At time $t_0$ determined by Eq.~6, the relative phase starts to increase linearly with the rate of $\Delta E /\hbar  \sim 2\pi / 4$ ps.
$t_0$ changes with $V_\text{ent}$ [see the blue line in Fig.~\ref{fig:supp_phase}(a)], but $t_0$ is independent of $V_\text{exit}$  [see the blue line in Fig.~\ref{fig:supp_phase}(b)].
These results confirm that the onset of the non-adiabatic excitation is determined by Eq.~6 of the main text. 

We discuss the probability $p$ of the non-adiabatic excitation in Eq.~1 in the main text. A non-trivial point is whether the excitation probability $p$ is sufficiently large in our experiment. Typically, $f_{\mathrm{in}}$ is 1 - 10 GHz~\cite{gib1, PTB-ulca1, NPL-NTT1, Zhao_pump}, which corresponds to the photon energy of $hf_{\mathrm{in}}\sim$ 4 - 40 $\mu$eV. On the other hand, $\Delta E$ is typically about 500 $\mu$eV~\cite{nttg1} or more. Based on a perturbation theory, one might expect that the non-adiabatic excitation does not occur, since $hf_\mathrm{in} \ll \Delta E$. However, our experiment is in a non-perturbative regime so that multiple photons can be absorbed by the electron in the QD, hence, the probability $p$ can be sufficiently large to have the coherent oscillations, as the numerical solution of the Schr\"{o}dinger equation indicates.

\section{Derivation of  the ejection probability in Eq. 2}
We derive the ejection probability $P_T$ in Eq. 2, based on the time-dependent scattering theory developed in Ref.~\cite{Sungguen1}.
 
 \begin{figure}[h]
\begin{center}
\includegraphics[width=0.7\textwidth]{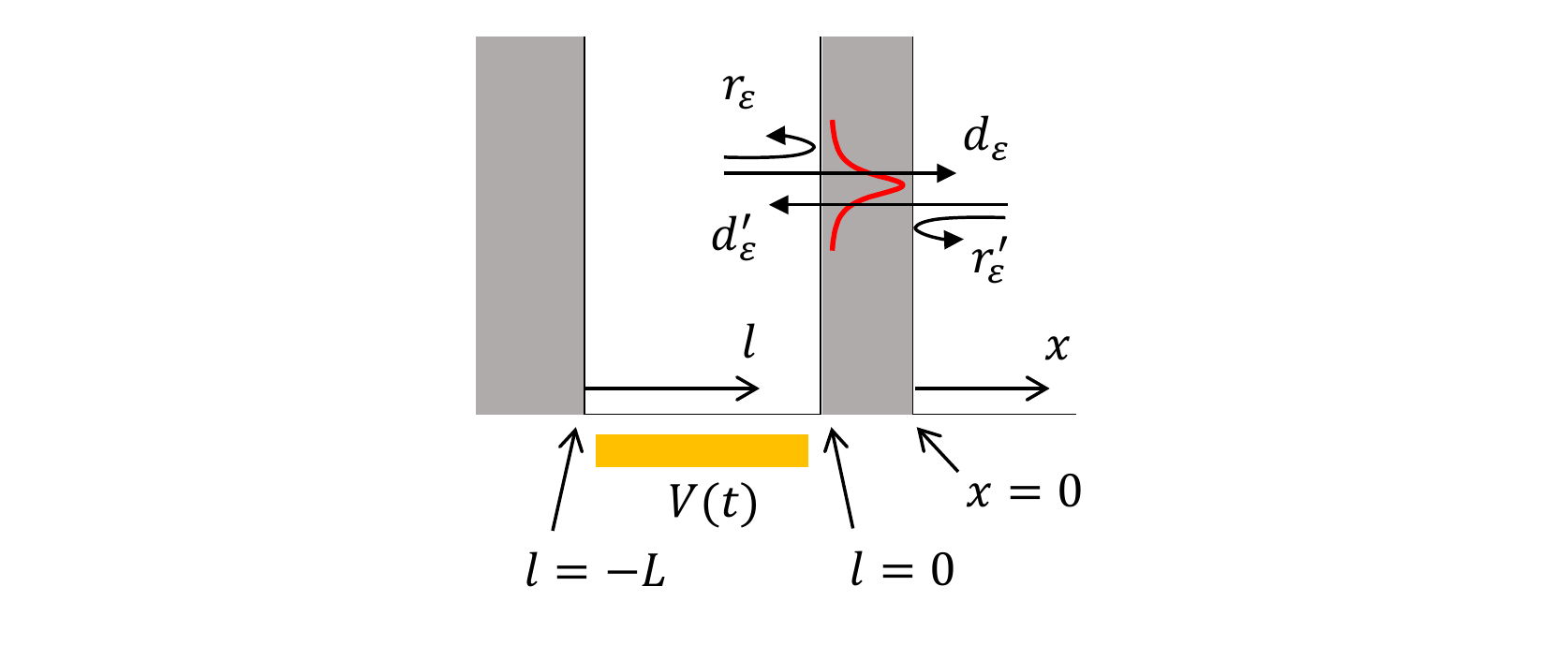}
 \end{center}
 \caption{Scattering model for electron ejection from the dynamic quantum dot (QD, the potential well of coordinate $l \in [-L,0]$) to the outside (the $x>0$ region) through the resonant level (depicted by the red Lorentzian peak).
The QD is formed between the left and right potential barriers (gray regions), and the resonant level exists in the right barrier.
The rise of the QD potential bottom is described by the time-dependent potential energy $V(t)$. 
The scattering of an electron plane wave with energy $\mathcal{E}$ by the resonant level is described by the reflection amplitudes ($r_\mathcal{E}$, $r'_\mathcal{E}$) and transmission amplitudes ($d_\mathcal{E}$, $d'_\mathcal{E}$) of the plane wave. }
 \label{fig_supp_eq2}
 \end{figure}

We explain the scattering model.
 The coherent time evolution of the wave packet inside the dynamic QD and its emission through the resonant level (see Fig. 1b in the main text) can be described by the model in Fig.~\ref{fig_supp_eq2}. The QD is simplified as the one-dimensional rectangular quantum well of coordinate $l \in [-L, 0]$. The rise of the potential bottom of the QD is described by the time-dependent potential energy $V(t)$. The resonant level is located inside the right barrier of the QD (between $l=0$ and $x=0$). 
 The transmission amplitude of the plane wave with energy $\mathcal{E}$ through the resonant level from the inside to outside (from the outside to the inside) of the QD is $d_\mathcal{E}$ ($d_\mathcal{E}'$), while its reflection amplitude inside (outside) the QD is $r_\mathcal{E}$  ($r_\mathcal{E}'$). The transmission amplitude $d_\mathcal{E}$ has a non-negligible value around the resonant level  $E_\text{res}$. This is described by 
 \begin{equation}
|d_\mathcal{E}|^2 = \frac{T_\text{max}}{1 + 4( \mathcal{E}-E_\text{res})^2/\Delta_\text{res}^2},
\end{equation} 
where $\Delta_\text{res}$ is the resonant-level broadening (see the red Lorentzian peak in Fig.~\ref{fig_supp_eq2}) and $T_\text{max}$ is the transmission probability at $\mathcal{E}=E_\text{res}$.
The energy dispersion relation inside and outside the QD is simplified as $\mathcal{E} = \hbar v k $, where $k$ is the momentum of the electron and $v$ is the electron velocity.
The simplifications introduced above are good approximations when the condition $\Delta_\text{res}/\dot{V} \ll \tau_\text{coh}$ (see the second inequality of Eq.~3 in the main text) is satisfied. The condition means that the time scale  $\Delta_\text{res}/\dot{V}$, within which the transmission amplitude of the plane wave with energy $\mathcal{E}$ through the resonant level is non-negligible, is much shorter than the period $\tau_\text{coh}$ of the coherent oscillations of the electron wave packet.
Since the scattering of the wave packet by the resonant level occurs within the short time scale $\Delta_\text{res}/\dot{V}$, it is well described by the simplified model. 
  
Using the scattering model, we solve the time evolution of the wave packet $\psi_0 = \psi_S (t_0)$ localised in the QD at initial time $t_0$ (see Eq.~1 in the main text). For the purpose, we obtain~\cite{Sungguen1} the time-dependent scattering state $\Psi_\mathcal{E} (t)$ of the electron plane wave incoming from $x=\infty$ to the QD with kinetic energy $\mathcal{E}$, 
and we write $\psi_0$ as the superposition $| \psi_0 \rangle = \int d \mathcal{E}\, a_{\mathcal{E}} |\Psi_\mathcal{E} (t_0) \rangle $ of those scattering states at time $t_0$.
Here, in the derivation of the scattering states, the effects of the time dependence $V(t)$ of the QD potential and the resonant level are taken into account. Then, the time evolution of the state $\psi_0$ at time $t> t_0$ is determined by $\int d \mathcal{E} \, a_{ \mathcal{E}} | \Psi_\mathcal{E} (t) \rangle $.

Using the spatial distribution of the time evolved wave function outside the QD, we derive the ejection probability $P_T$. Below, we provide the details of the derivation of $P_T$.

We first obtain the scattering state $|\Psi_\mathcal{E}(t) \rangle$. This state results from the scattering, by the resonant level, of a plane wave $e^{-i \mathcal{E}(t+x/v)}$ of energy $\mathcal{E}$ incoming from $x = \infty$ outside the QD.  
This state is written as a superposition of states of energy different from $\mathcal{E}$ because the QD potential has the time dependence $V(t)$. 
To treat the time dependence, we apply a gauge transformation of the QD  potential bottom $\Phi$ and the vector potential $\mathbf{A}$,
\begin{equation}
  \begin{aligned}
    \Phi = V(t)  \quad   \rightarrow \quad & \Phi -\frac{\partial \Lambda}{\partial t}=0\\
    \mathbf{A} =0 \quad  \rightarrow \quad & \mathbf{A} + \mathbf{\nabla} \Lambda =-\delta (l) \int_{-\infty}^t V(u)du \\
 &   \Lambda = \Theta(-l) \int_{-\infty}^t V(u) \, du,
  \end{aligned}  \label{eq:GT}
\end{equation}
where $\Theta(l)$ and $\delta (l)$ are the step function and the delta function, respectively; $\Theta(l) = 1$ for $l>0$ and 0 for $l<0$; hereafter we use $\hbar \equiv 1 $ in this section.
After the gauge transformation, the QD potential becomes time independent. Instead, the electron wave function inside the QD gains the phase factor
$e^{i\phi(t)}  \equiv  e^{i \int_{-\infty}^{t} V(u) \, du}$. 
In terms of the phase factor and the scattering amplitudes $d_\mathcal{E}$, $r_\mathcal{E}$, $d'_\mathcal{E}$, $r'_\mathcal{E}$, we obtain $|\Psi_\mathcal{E} \rangle $ (see the steps in Eqs. S11-S17 in Ref.~11. The state form is of Fabry-Perot type. Inside the QD, $|\Psi_\mathcal{E} \rangle $ is found as 
\begin{align}
\label{eq:Psi_in_dot}
  \langle l | \Psi_{\mathcal{E}}(t) \rangle = & e^{i \phi(t+l/v)}   d'_{\mathcal{E}}   e^{-i \mathcal{E} (t+l/v)}  \nonumber \\
&  +\sum_{M=1}^\infty  e^{i\phi(t+l/v -M\tau_\text{coh})} e^{iM \mathcal{E} \tau_\text{coh}}(-1)^M
  \left[\Pi_{m=1}^M r_{\mathcal{E}+ V(t+l/v -(M-m)\tau_\text{coh} - V(t+l/v -M\tau_\text{coh})) } \right] d'_\mathcal{E} e^{-i \mathcal{E} (t+l/v)} \nonumber \\ 
& + [\text{the same term but with the replacement of } l \rightarrow (-l +v \tau_\text{coh})].
\end{align}
Notice that the phase factors $e^{i\phi(t)}$, resulting from the time dependence of the QD potential, are attached to the wave function inside the QD.
Outside the QD, $|\Psi_\mathcal{E} \rangle $ is obtained as
\begin{equation}
\label{eq:Psi_out_dot}
\begin{aligned}
&\langle x | \Psi_{ \mathcal{E}} (t) \rangle  = 
 e^{-i \mathcal{E} (t +x/v) }+ r'_{\mathcal{E}} e^{-i \mathcal{E} t_r}  
 \\
&\quad + \sum_{M=1}^{\infty} d_{\mathcal{E}+V(t_r) -V(t_r-M\tau_\text{coh}) }  e^{-i\phi(t_r)+i\phi(t_r-M\tau_\text{coh})} e^{i M  \mathcal{E} \tau_\text{coh}} (-1)^M 
\left[\Pi_{m=1}^{M-1} r_{ \mathcal{E} +  V(t_r - (M-m)\tau_\text{coh}) - V(t_r -M\tau_\text{coh}))} \right] d'_\mathcal{E} e^{-i \mathcal{E} t_r}, 
\end{aligned}
\end{equation}
where $t_r \equiv t - x/v$.
Each term of index $M$ in Eq.~\ref{eq:Psi_out_dot} describes the process that the incident electron of energy $\mathcal{E}$ enters the QD at time $t_r - M \tau_\text{coh}$,  travels the distance $2L$ of the QD (from its right end $l=0$ to the left end $l=-L$ and then to the right end)  $M$ times, and then escapes from the QD at time $t_r$. In the $M = 1$ term we use $\Pi_{M=1}^0 \equiv 1 $ instead of 0, for brevity. Note that Eq.~\ref{eq:Psi_out_dot} corresponds to Eq.~S17 in Ref.~\cite{Sungguen1} except the factor $(-1)^M$, which comes from the boundary condition that the wave function vanishes at the left end $l=-L$ of the QD.

Now we write the initial localised wave packet $\psi_0$ at time $t_0$ as the superposition $| \psi_0 \rangle = \int d \mathcal{E}\, a_{\mathcal{E}} |\Psi_\mathcal{E} (t_0) \rangle $ of the scattering states, and obtain the expansion coefficient $a_\mathcal{E}$.   
We first simplify the Eq.~\ref{eq:Psi_in_dot} at time $t_0$ by choosing $V(t) = V(t_0)$ for time $t < t_0$ (this choice does not affect the time evolution of the packet $\psi_0$), 
\begin{equation}
  \langle l | \Psi_{ \mathcal{E}} ( t_0) \rangle
  = e^{i \phi(t_0)} e^{-i \mathcal{E} t_0} 
   e^{-i (\mathcal{E} -V(t_0)) \frac{l}{v} } \left\{ d'_\mathcal{E} + \frac{d'_\mathcal{E}}{1 + r_\mathcal{E} e^{i (\mathcal{E}-V(t_0))\tau_\text{coh} } } \right\} - [\textrm{the same term but with } l \rightarrow (-l + v \tau_\text{coh})]. 
\end{equation}
Because at time $t_0$ the packet in the QD is localised inside the QD (namely the electron does not exist outside the QD),  $d'_\mathcal{E} \rightarrow 0$ and $r_\mathcal{E} \rightarrow -1$ are satisfied. Hence, the first term vanishes and
the second term is nonvanishing near the energy $\mathcal{E}$ that satisfies the resonant condition of $\mathcal{E} = V(t_0) + E_n $, where $E_n \equiv 2n \pi/ \tau_\text{coh}$ is the energy quantisation of the QD, $n=1,2,\cdots$, and $V(t_0) + E_n$'s are the energy levels of the QD at time $t_0$. The second term is well approximated
by the sum of Lorentzian peaks at the resonant energies,
\begin{equation} \label{eq:Psi_in_QD_t0}
  \langle l | \Psi_{ \mathcal{E}} ( t_0) \rangle \simeq e^{i \phi(t_0)}   
  \sum_{n=1}^\infty 2 e^{-i (E_n+V(t_0)) t_0} \sin \frac{E_n l}{v} \frac{d'_\mathcal{E}/\tau_\text{coh}}{ \mathcal{E} - V(t_0) - E_n +i \frac{|d_\mathcal{E}|^2}{2\tau_\text{coh}} }.
\end{equation}
Using Eq.~\ref{eq:Psi_in_QD_t0}, the coefficient in $| \psi_0 \rangle = \int d \mathcal{E}\, a_{\mathcal{E}} |\Psi_\mathcal{E} (t_0) \rangle $ is obtained as
\begin{equation}
\label{eq:aE}
  a_{\mathcal{E}} = -\sum_{n'=1}^\infty \left( \frac{d'_\mathcal{E}/\tau_\text{coh}}{ \mathcal{E} - V(t_0) - E_{n'} +i \frac{|d_\mathcal{E}|^2}{2\tau_\text{coh}} }\right)^*
  e^{i \mathcal{E}t_0} \frac{\tau_\text{coh}}{4\pi} \sqrt{\frac{2}{L}} \langle n'_\text{QD}| \psi_0 \rangle,
\end{equation}
where
$| n'_\text{QD} \rangle$ is the $n'$th eigenstate of the QD at time $t_0$, satisfying $ \langle l | n'_\text{QD} \rangle = -\sqrt{2/L}  \sin (E_{n'} l /v) $.
In the derivation of Eq.~\ref{eq:aE}, we used the identity of
\[
  \sum_{n=1,n'=1}^{\infty} \left( \frac{d'_\mathcal{E}/\tau_\text{coh}}{ \mathcal{E} - V(t_0) - E_{n'} +i \frac{|d_\mathcal{E}|^2}{2\tau_\text{coh}} }\right)^* 
 \frac{d'_\mathcal{E}/\tau_\text{coh}}{ \mathcal{E} - V(t_0) - E_{n} +i \frac{|d_\mathcal{E}|^2}{2\tau_\text{coh}} } f_{n n'}
 \stackrel{d_\mathcal{E}, d'_\mathcal{E} \rightarrow 0 }{= } \frac{2\pi}{\tau_\text{coh}} \sum_{n=1}^\infty \delta(\mathcal{E} - V(t_0) -E_n) f_{nn},
\]
where $f_{nn'} $ is any arbitrary function of $n$ and $n'$. 

Next, we obtain the emitted part of the time evolved wave function $\psi(x,t) = \int d \mathcal{E}  a_\mathcal{E} \langle x | \Psi_{ \mathcal{E}} ( t) \rangle$ at $t>t_0$. 
For the purpose,  we multiply Eq.~\ref{eq:aE} and Eq.~\ref{eq:Psi_out_dot} and integrate it with respect to $\mathcal{E}$. Using the Fourier transform of the Lorentzian function, we obtain the emitted part
\begin{equation}
\label{eq:psi_out}
\psi(x,t) =  \sqrt{\frac{2}{L}}   \sum_{n,m=1}^\infty  \langle n_\text{QD}|\psi_0\rangle d_{E_n +V(t_r)}  
\left[ \Pi_{m'= 0 }^{m-1} (-1)r_{E_n + V(t_r -(m -m') \tau_\text{coh} )} \right] e^{-i \int_{t_0}^{t_r} (E_n +V(t')) dt' }
    \zeta_m (t_r) .
\end{equation}
Here $ \zeta_m(t_r) $  is 1 for $ t_r \in [t_0 +m \tau_\text{coh}, t_0 + (m+1)\tau_\text{coh} ] $ and 0 otherwise. Note that this corresponds to Eq.~S19 in Ref.~11 except the sign factor $(-1)$ that comes from the boundary condition at the left end $l=-L$ of the QD. The physical meaning of Eq.~\ref{eq:psi_out} is as follows. Each $(n,m)$ term describes that an electron occupies the $n$th QD level at time $t_0$ and then is emitted after $m$ oscillations. 
It gains the dynamical phase $ \int_{t_0}^{t_r} (E_n + V(t')) dt'$. The amplitude of the emission is determined by $m$ reflections at the energies $E_n+V(t_r-m \tau_\text{coh}), E_n+V(t_r-(m-1) \tau_\text{coh})$ $, \cdots, E_n+V(t_r -\tau_\text{coh}) $  and the final transmission through the resonant level at the energy $E_n+V(t_r)$.  $ \langle n_\text{QD}|\psi_0\rangle$ is the weight that an electron occupies the $n$th level of the QD at time  $t_0$.

Finally, we obtain the ejection probability $P_T$ of $\psi_0$ through the resonant level, with applying the wave packet form of $\psi_0$ in Eq.~1 in the main text. For the purpose, we plug $\langle 1_\text{QD} | \psi_0 \rangle = \sqrt{1-p}$ and $\langle 2_\text{QD} | \psi_0 \rangle = \sqrt{p}e^{i \theta} $ into Eq.~\ref{eq:psi_out} and then compute $P_T = v \int_{t_0}^\infty dt |\psi(0,t)|^2$.  In the computation, we use approximations applicable under the condition of $\Delta E \lesssim \Delta_\text{res} \lesssim \tau_\text{coh} \dot{V}$. Due to the second inequality, the reflection amplitudes in Eq.~\ref{eq:psi_out} is approximately $-1$ for non-negligible $d_{E_n +V(t)}$. Due to the first inequality, the transmission amplitude is approximated as $d_{(E_1+E_2)/2 +V(t) }$. Then, $P_T$ is simplified as
\begin{equation}
\label{eq:before-PT}
  P_T \simeq   \frac{2 v}{L}  \int_{t_0}^\infty dt\, \left| d_{(E_1+E_2)/2+V(t)} \right|^2  \left| \sqrt{1-p} + \sqrt{p} e^{-i(E_2 -E_1)(t-t_0)+i\theta}  \right|^2 .
\end{equation}
Under the condition of the second inequality of  $\Delta E \lesssim \Delta_\text{res} \lesssim \tau_\text{coh} \dot{V} $, the time in the second absolute square in Eq.~\ref{eq:before-PT} can be approximated as the time when the mean energy of the wave packet and the resonant level is aligned.  We finally obtain Eq.~2 of the main text, after integrating the transmission probability in time, $\int_{t_0}^\infty dt \, \big| d_{(E_1+E_2)/2+V(t)} \big|^2 = (\pi/2) T_\text{max} \Delta_\text{res}/\dot{V}$. 

 \section{Signature of non-adiabatic excitation}
 \begin{figure}[h]
\begin{center}
\includegraphics[pagebox=artbox]{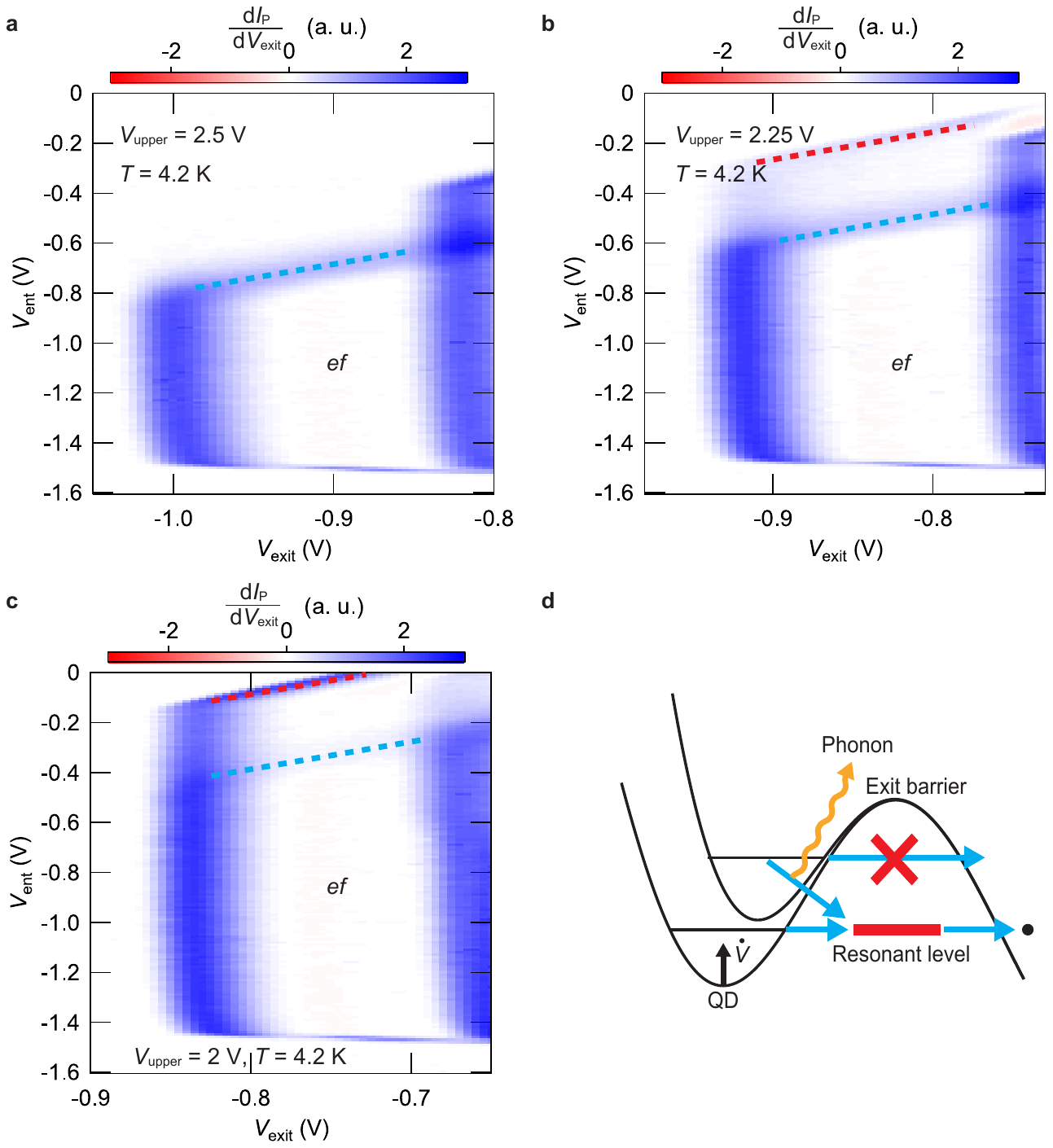}
 \end{center}
 \caption{\bf a-c\rm, First derivative of $I_{\mathrm{P}}$ with respect to $V_{\mathrm{exit}}$ as a function of $V_{\mathrm{ent}}$ and $V_{\mathrm{exit}}$ at $f_{\mathrm{in}}=50$ MHz and $T=4.2$ K, where $P=9$ dBm and $V_{\mathrm{upper}}$ are $2.5$ V (\bf a\rm), $2.25$ V (\bf b\rm), and $2$ V (\bf c\rm). \bf d\rm, Schematic potential diagram showing the resonant and inelastic tunneling through the resonant level in the exit barrier. $\dot{V}$ is the rising speed of the QD bottom. The inelastic tunneling occurs with a phonon emission.}
 \label{f6}
 \end{figure}
We discuss the experimental signature of the non-adiabatic excitation in our device. At first, we start from the low-frequency 50-MHz data, which should be in the adiabatic regime. Figure \ref{f6}a shows a $\mathrm{d}I_{\mathrm{P}}/\mathrm{d}V_{\mathrm{exit}}$ map as a function of $V_{\mathrm{ent}}$ and $V_{\mathrm{exit}}$ with $V_{\mathrm{upper}}=2.5$ V. In this map, there is only an ejection line related to the direct tunneling through the exit barrier (blue dashed line). However, with decreasing $V_{\mathrm{upper}}$ (Figs. \ref{f6}b and \ref{f6}c), an additional line appears (red dashed line) and the ejection line almost disappears at $V_{\mathrm{upper}}=2$ V. The additional current flow indicates that there is another current path through a resonant level, which is probably due to the interface trap level in the exit barrier\cite{trap1, trap2, rossi_trap}. We refer the red dashed line to as the trap-ejection line. Between the ejection and trap-ejection lines, the direct tunneling through the exit barrier is suppressed but the resonant tunneling through the resonant level during the rise of the QD energy level and the inelastic tunneling with phonon emission after the QD energy level is higher than the resonant level would occur\cite{FujiPh} (Fig. \ref{f6}d). These current flows strongly depend on the coupling between the QD energy and resonant levels. When we increase $V_{\mathrm{upper}}$, the central part of the QD mainly lowers because of the screening by the lower gates (see Fig. 2 in the main text), resulting in a stronger confinement of the QD. This would reduce the coupling between the QD energy and resonant levels, leading to the decrease in the current through the resonant level.

\begin{figure}[h]
\begin{center}
\includegraphics[pagebox=artbox]{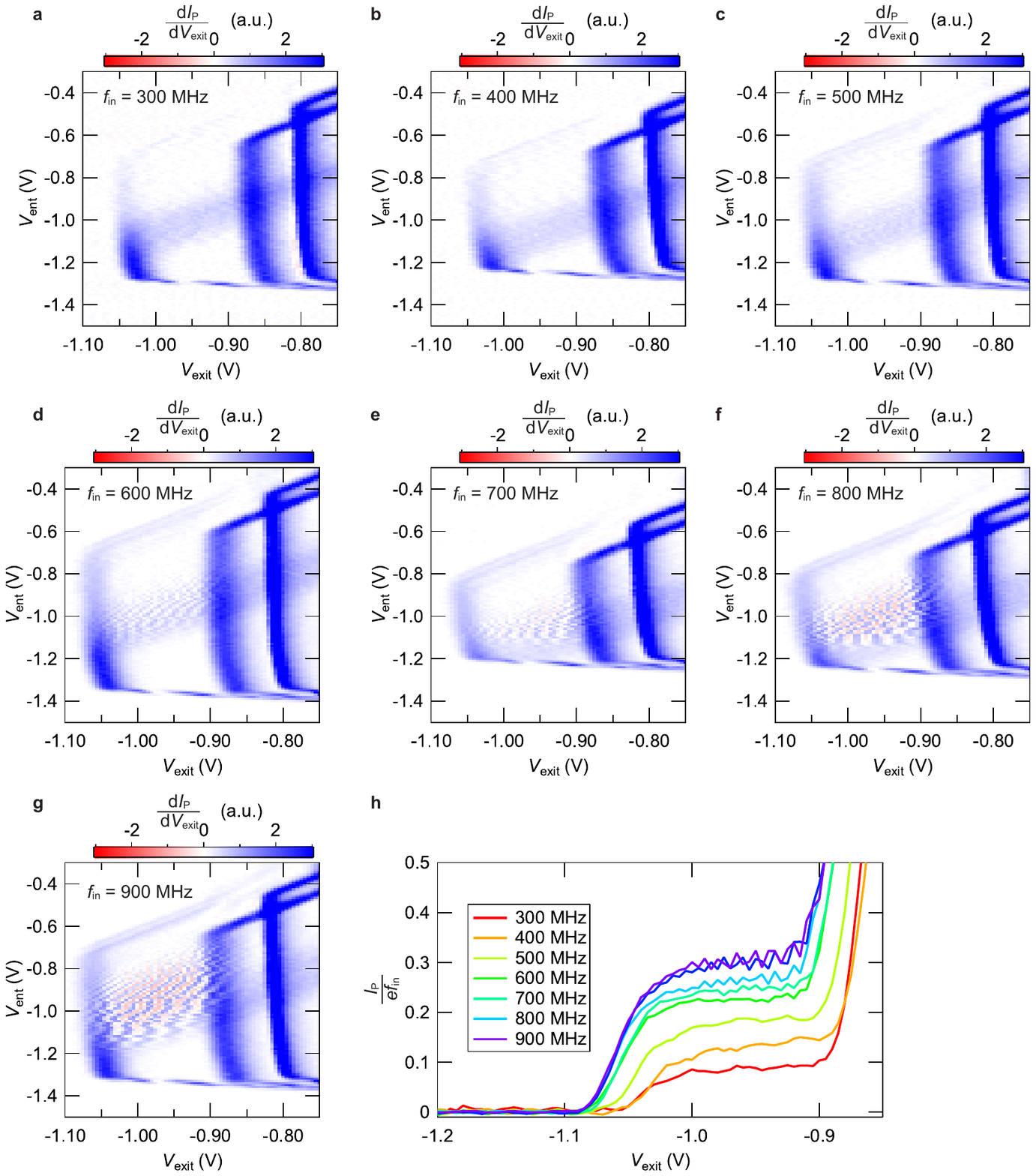}
 \end{center}
 \caption{\bf a-g\rm, First derivative of $I_{\mathrm{P}}$ with respect to $V_{\mathrm{exit}}$ as a function of $V_{\mathrm{ent}}$ and $V_{\mathrm{exit}}$ at $T=4.2$ K, where $V_{\mathrm{upper}}=2.5$ V, $P=9$ dBm, and $f_{\mathrm{in}}$ are 300 MHz (\bf a\rm), 400 MHz (\bf b\rm), 500 MHz (\bf c\rm), 600 MHz (\bf d\rm), 700 MHz (\bf e\rm), 800 MHz (\bf f\rm), 900 MHz (\bf g\rm). \bf h\rm, $I_{\mathrm{P}}$ normalised by $ef_{\mathrm{in}}$ as a function of $V_{\mathrm{exit}}$ at $T=4.2$ K and $f_{\mathrm{in}}$ of 300 to 900 MHz, where $V_{\mathrm{upper}}=2.5$ V and $P=9$ dBm. We select $V_{\mathrm{ent}}$ such as $V_{\mathrm{ent}}=V_{\mathrm{ent}}^{\mathrm{cross}}-0.05$ (V), where $V_{\mathrm{ent}}^{\mathrm{cross}}$ at the crossing point of the trap-ejection line and the capture line from 0 to $ef_{\mathrm{in}}$}.
 \label{f7}
 \end{figure}

Then, we fix $V_{\mathrm{upper}}=2.5$ V, at which the tunneling rate through the resonant level from the ground state of the QD is low. With increasing $f_{\mathrm{in}}$, the trap-ejection line appears and becomes clear (Figs. \ref{f7}a - \ref{f7}g). This indicates that the excited states are populated, because the wave functions of the excited states have peaks closer to the edge of the QD than that of the ground state and the coupling between the QD energy and resonant levels is stronger. In addition, the current level normalised by $ef_{\mathrm{in}}$ between the ejection and trap-ejection lines increases with increasing $f_{\mathrm{in}}$ in spite of the decreased time duration during which the QD energy level is higher than the resonant level (Fig. \ref{f7}h). This indicates that the excitation probability increases with increasing $f_{\mathrm{in}}$, which is consistent with the previous report\cite{kataoka1}. Note that the ejection line becomes broad when the excitation occurs, which would be a signature of the inelastic tunneling\cite{FujiPh}.
 
In addition, the current oscillations appear at the ejection line from the 600-MHz data and the oscillations become clearer at higher $f_{\mathrm{in}}$. Since the period of the current oscillations is large enough compared with the voltage resolution, the decrease of the contrast with decreasing $f_{\mathrm{in}}$ would be related to the decoherence. Actually, we observe low contrast of the current oscillations near the trap-ejection line at 1 GHz (around the blue arrow in Fig.~4a in the main text), but the 2-GHz data does not have a similar decrease of the contrast (around the blue arrow in Fig.~4b in the main text). Since $t_{1}-t_{0}$ is maximal at the trap-ejection line, the decrease of the contrast might be due to decoherence. If so, the decoherence time is 0.1 - 1 ns, which is of a similar order of magnitude as the decoherence time of typical coherent charge oscillations\cite{ntte, Petersson1, DKim1}. Note that the spacing between the loading and trap-ejection lines are different at different $f_{\mathrm{in}}$ because the cross talk of the high-frequency signal is different at different $f_{\mathrm{in}}$.
\clearpage

\section{Estimation of $t_{1}$}
The final time $t_{1}$ of the coherent time evolution can be the time when the mean QD energy level is aligned with the resonant level. We approximately use $E_{1}^{\mathrm{qd}}$ instead of the mean energy: $E_{1}^{\mathrm{qd}}(t_{1})=E_{\mathrm{res}}(t_{1})$. The approximation is valid with the condition of Eq. 3 in the main text. Since $V_{\mathrm{ac}}(t)=V_{\mathrm{amp}}\mathrm{cos}\left (2\pi f_{\mathrm{in}}t\right)$, $E_{1}^{\mathrm{qd}}$ can be written as (see Eq. \ref{Edot})
\begin{equation}
E_{1}^{\mathrm{qd}}(t_{1})=E^{\mathrm{qd}}_{\mathrm{off}}-\alpha _{\mathrm{ent\tiny \_\normalsize  QD}}V_{\mathrm{amp}}\mathrm{cos}\left (2\pi f_{\mathrm{in}}t_{1}\right) \label{e1qd}.
\end{equation}
To estimate $E_{\mathrm{res}}$, we assume that the gate dependence of $E_{\mathrm{res}}$ can be the same as $U^{\mathrm{exit}}(t)$ because the trap-ejection line is parallel with the ejection line. Then, $E_{\mathrm{res}}$ can be written as (see Eq. \ref{exitB})
\begin{equation}
E_{\mathrm{res}}(t_{1})=U_{\mathrm{off}}^{\mathrm{exit}}- \alpha _{\mathrm{ent\tiny \_\normalsize  exitBarrier}}V_{\mathrm{amp}}\mathrm{cos}\left (2\pi f_{\mathrm{in}}t_{1}\right). \label{eres}
\end{equation}
From Eqs. \ref{U2}, \ref{QDoff}, \ref{e1qd}, and \ref{eres}, $E_{1}^{\mathrm{qd}}(t_{1})=E_{\mathrm{res}}(t_{1})$ leads to
\begin{equation}
t_{1}=\frac{1}{2\pi f_{\mathrm{in}}}\mathrm{cos}^{-1}\left[ \frac{1}{V_{\mathrm{amp}}}\left(-V_{\mathrm{ent}}+\frac{\alpha ^{\mathrm{E}}_{\mathrm{exit}}}{\alpha^{\mathrm{E}}_{\mathrm{ent}}}V_{\mathrm{exit}}\right ) \right],
\end{equation}
where we neglect the gate-independent constant terms for simplicity.

\section{Additional Discussion of Figs 4 and 5}
There is some mismatch between Figs.~4 and 5 in the main text, although the calculation reproduces the main features of the experimental results. For example, the mismatch of the voltage axis values is due to ignorance of some gate-independent constants in the derivation of $t_1$. In addition, it is difficult to reproduce the amplitude of the current oscillations, because it depends on many parameters ($\Delta _{\mathrm{res}}$, $p$, $\Gamma _{\mathrm{L(R)}}$, decoherence) and because $\Delta E$ and the alpha factors can weakly change dynamically. 

\section{Evaluation of Eq. 3}
For the evaluation of Eq. 3 in the main text, we roughly estimate $\dot{V}$. Since the gate-QD coupling changes in the capture and ejection stages (see Table \ref{tab:price}), we use an averaged value  $\alpha_{\mathrm{ave}} = (\alpha _{\mathrm{ent\tiny \_\normalsize QD}}^{\mathrm{C}}+\alpha _{\mathrm{ent\tiny \_\normalsize QD}}^{\mathrm{E}})/2 \sim 0.27$ eV/V. Then, we estimate the average value of $\dot{V}$ during the first half of the pumping,
$\dot{V}\sim 4f_{\mathrm{in}}\alpha_{\mathrm{ave}}V_{\mathrm{amp}}\sim 1.54$ eV/ns at 1 GHz. Since $\Delta E\sim 1$ meV, $\tau_{\mathrm{coh}}\dot{V}=h\dot{V}/\Delta E \sim 6.4$ meV. 

\section{Landau-Zener-St\mbox{\boldmath $\ddot{\mathrm{U}}$}ckelberg interference}
Since the QD energy and resonant levels can be considered as a double-QD system, the Landau-Zener-St$\ddot{\mathrm{u}}$ckelberg (LZS) interference\cite{LZS1, LZSSi1} could be a candidate of the origin of the current oscillations. However, this is not the case as explained below. 

\begin{figure}
\begin{center}
\includegraphics[pagebox=artbox]{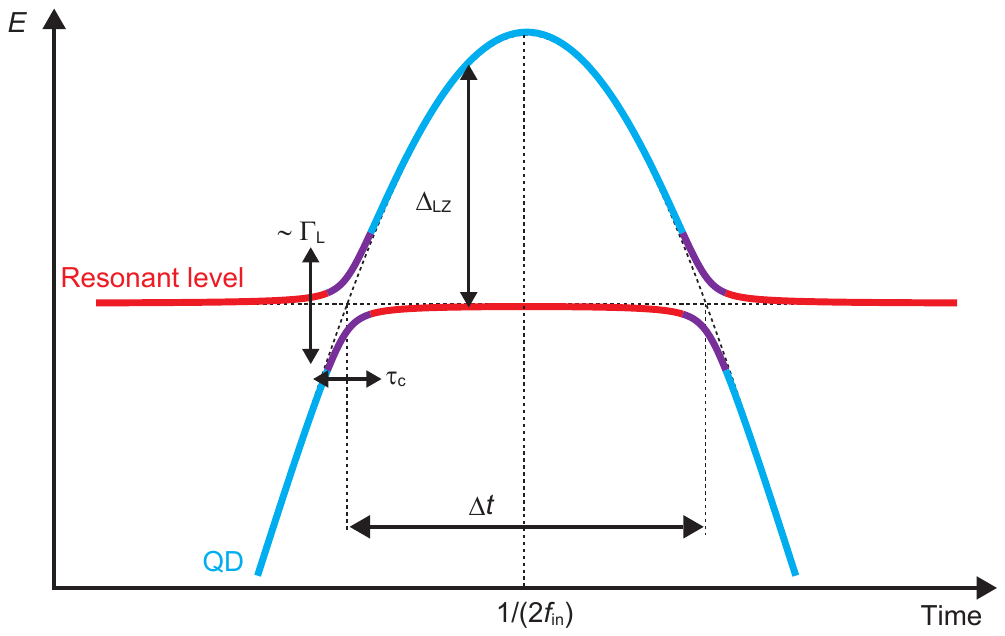}
\end{center}
 \caption{Schematic energy diagram as a function of time. The red and blue lines correspond to the resonant and QD energy levels, respectively. The purple lines indicate the coupling regime of these two levels with coupling energy $\Gamma _{\mathrm{L}}$ and coupling time $\tau _{\mathrm{c}}$. $\Delta _{\mathrm{LZ}}$ is the energy difference between the QD energy and resonant levels. $\Delta t$ is the time duration between the two crossing points.}
 \label{f8}
 \end{figure}

Figure \ref{f8} shows a schematic energy diagram of the QD energy (blue line) and resonant (red line) levels as a function of time. When the QD energy level is close to the resonant level, an avoided crossing of them occurs (purple line)\cite{dqd1}, which depends on the coupling energy $\Gamma _{\mathrm{L}}$ between the two levels. Since the QD energy level passes the resonant level twice, the LZS interference could occur. When the final state is the resonant (QD energy) level, there should be (no) current flows. The interference period depends on the accumulated phase $\phi _{\mathrm{LZ}}$ after the QD energy level crosses the resonant level, which is written as
\begin{equation}
\phi _{\mathrm{LZ}}=\frac{2}{\hbar}\int _{t^{\mathrm{1st}}}^{\frac{1}{2f_{\mathrm{in}}}}\Delta_{\mathrm{LZ}}dt,
\end{equation}
where $\Delta _{\mathrm{LZ}}$ is the energy difference between the QD energy ($E_{1}^{\mathrm{qd}}$) and resonant ($E_{\mathrm{res}}$) levels and $t^{\mathrm{1st}}$ is the time when the QD energy level crosses the resonant level for the first time. $\Delta _{\mathrm{LZ}}$ has a correction term due to the avoided crossing, which is order of $\Gamma _{\mathrm{L}}$ in time duration $\tau _{\mathrm{c}}$ (see the purple line). The integral of the correction term is on the order of $\Gamma _{\mathrm{L}}\tau _{\mathrm{c}}/\hbar$. When the QD movement is neither adiabatic nor sudden ones, $\Gamma _{\mathrm{L}}\tau _{\mathrm{c}}/\hbar$ is on the order of 1. Then, when $V_{\mathrm{ent}}$ changes with an amount of $\Delta V_{\mathrm{ent}}$, the change in $\phi _{\mathrm{LZ}}$ can be calculated as
\begin{equation}
\Delta \phi _{\mathrm{LZ}}\sim \frac{\alpha _{\mathrm{ent\tiny \_\normalsize QD}}-\alpha _{\mathrm{ent\tiny \_\normalsize exitBarrier}}}{\hbar}\Delta V_{\mathrm{ent}}\Delta t + \mathrm{(order\ of\ 1)},
\end{equation}
where $\Delta t$ is the time between the first and second energy crossings (see Fig. \ref{f8}). To observe one oscillation in the pump map, we need the change in the entrance gate voltage as
\begin{equation}
\Delta V_{\mathrm{ent}} \sim \frac{(2\pi+[\mathrm{order\ of\ 1}])\hbar}{(\alpha _{\mathrm{ent\tiny \_\normalsize QD}}-\alpha _{\mathrm{ent\tiny \_\normalsize exitBarrier}})\Delta t}.
\end{equation}
When we increase $V_{\mathrm{ent}}$, the QD energy level is lowered with respect to the resonant level, resulting in smaller $\Delta t$. This indicates that the oscillation period ($\Delta V_{\mathrm{ent}}$) increases with increasing $V_{\mathrm{ent}}$. This is opposite to the experimental observation (see Fig.~4b in the main text). Thus, we conclude that the LZS interference is not the origin of the current oscillations. The reason why we do not observe the LZS interference might be that the following condition is not satisfied: $\Gamma_{\mathrm{L}}<\hbar/\Delta t <\Gamma_{\mathrm{R}}$.

\section{Protocol for detecting fast wave-packet dynamics in a cavity}
 \begin{figure}[h]
\begin{center}
\includegraphics[width=0.7\textwidth]{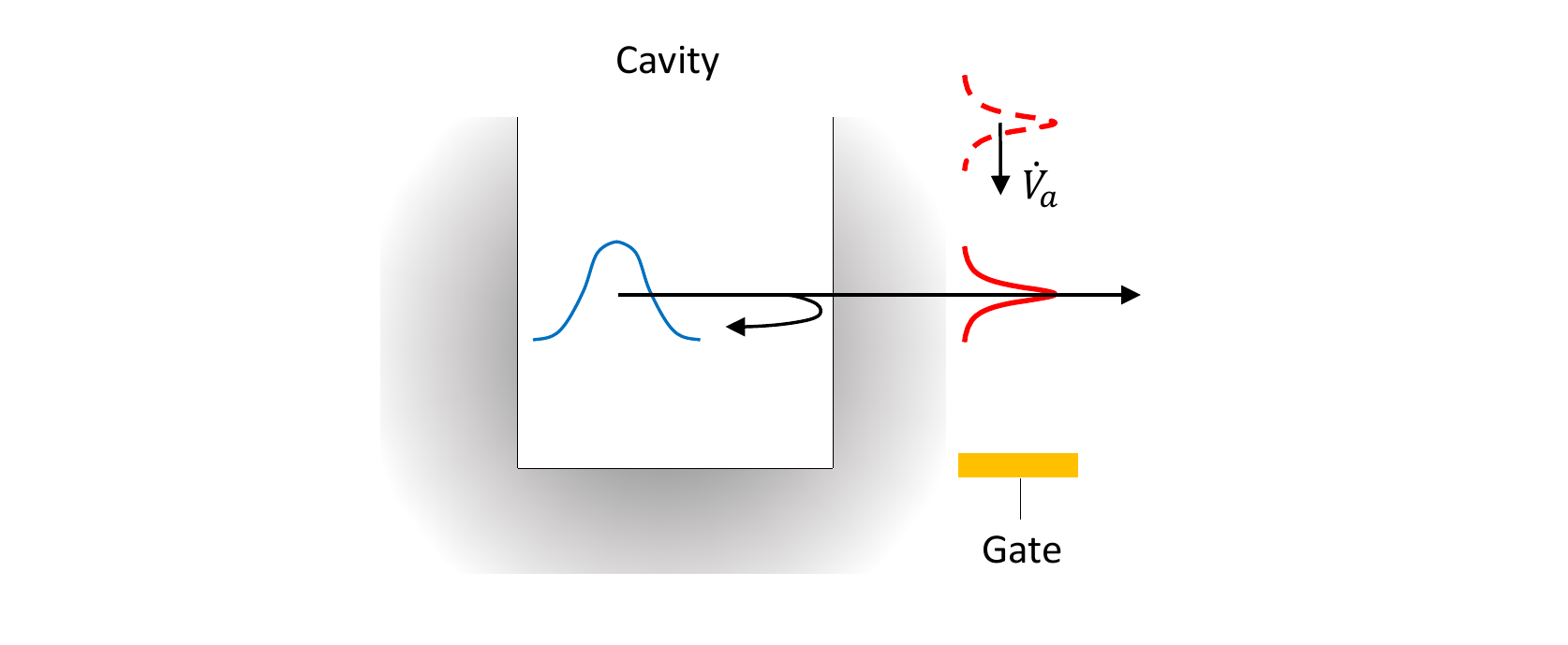}
 \end{center}
 \caption{Protocol for detecting fast wave-packet dynamics in a cavity. The cavity hosts coherent oscillations (depicted by arrows) of a wave packet (a blue peak) of a particle. The oscillations are detected by the particle current through a time-dependent resonant level (red Lorentzian peak), formed by an artificial atom such as a quantum dot. The shift (vertical arrow) of the resonant level is driven by a local gate (orange rectangle). 
}
 \label{fig_supp_protocol}
 \end{figure}

Here, we discuss an implication of our results. Our findings suggest a general protocol for detecting fast dynamics of a wave packet in a cavity. 

Coherent wave-packet oscillations can generally occur when a particle (such as an electron) is confined in a cavity (such as a quantum dot) and driven by AC voltages in a non-adiabatic fashion. This can happen in quantum nanodevices operated for many purposes. 
The motion of such wave packets is typically much faster than the range directly measurable with currently available bandwidth. Based on our findings in the main text, we propose a protocol for detecting such fast coherent wave-packet oscillations using a resonant level formed in an artificial atom such as a quantum dot.

Figure~\ref{fig_supp_protocol} shows a setup for the protocol, which consists of a cavity hosting coherent wave-packet spatial oscillations of a particle and an artificial atom having a resonant energy level.
Initially, the resonant energy level (see the red dashed Lorentzian peak in the figure) is much higher than the energy of the particle.
Then the energy of the resonant level decreases with rate $\dot{V}_a$ and reaches the value far below the energy of the particle (below at least by the energy uncertainty of the particle wave packet).
When the resonant level becomes aligned with the energy of the particle, the particle can transmit through the resonant level to move out of the cavity.
The transmission probability is large (small) if the particle wave packet is located at the right (left) side of the cavity.
Hence, by measuring the current of the particle outside the cavity, one can get the information of the coherent spatial oscillations of the particle.
Note that an amount of the current large enough for the detection can be collected by repeating the above process from the initialization of the wave-packet oscillations to the time-dependent change of the resonant level.

This protocol corresponds to the measurement of the coherent spatial oscillations of an electron wave packet shown in the main text; the resonant level is artificially generated here in the protocol.
The time resolution of the protocol is $\Delta_\text{res}/\dot{V}_a$, according to the second inequality of Eq.~3 in the main text. The time resolution can be much larger than the currently available experimental bandwidth, when sufficiently large $\dot{V}_a$ is applied. 
For example, using the energy broadening $\Delta_\text{res} \sim 1 $ meV of the resonant level and $\dot{V}_a \sim $ 1 eV/ns (which are within experimental reach, as shown in the main text), 
one can achieve the resolution of $\Delta_\text{res} /\dot{V}_a = 1$ ps.
The resolution is equivalent with 1 THz,  which is far in excess of currently achievable bandwidth of 10 GHz.